\documentclass[twoside,12pt]{article}

\usepackage{epsfig,array}
\usepackage{amssymb}
\usepackage{amsmath}
\usepackage[english]{babel}
\usepackage{afterpage}
\def\setb@se#1{\baselineskip=#1 \normalbaselineskip=#1}
\setb@se{14pt}
\begin{document}
\def\theequation{\arabic{equation}}
\def\Section#1{\setcounter{equation}{0}\section{#1}}
\renewcommand{\thefootnote}{\fnsymbol{footnote}}
\numberwithin{equation}{section}
\pagestyle{empty}
\pagenumbering{arabic}

\begin{flushright}
ZU-TH 13/97\\
\end{flushright}

\begin{center}
{\Large   Generalized Heavy Baryon Chiral Perturbation
Theory\footnote{Partially supported by Schweizerischer
Nationalfonds.}}\\[15mm]
Robert Baur\footnote{e-mail: \makeatletter rbaur@physik.unizh.ch
\makeatother}  and Joachim Kambor\footnote{
e-mail: \makeatletter kambor@physik.unizh.ch \makeatother}\\[2mm]
Institut f\"ur Theoretische Physik, Universit\"at Z\"urich\\
CH-8057 Z\"urich, Switzerland\\[18mm]
{\large {\bf Abstract}}\\
\end{center}
Standard $SU(2)$ Heavy Baryon Chiral Perturbation Theory is extended
in order to include the case of small or even vanishing quark
condensate. The effective lagrangian is given to
${\cal O}(p^2)$ in its most general form and to ${\cal O}(p^3)$ in the
scalar sector. A method is developed to efficiently
construct the relativistic baryonic effective lagrangian for chiral
SU(2) to all orders in the chiral expansion.
As a first application, mass- and wave-function
renormalization as well as the scalar form factor of the nucleon is
calculated to ${\cal O}(p^3)$. The result is compared to a dispersive
analysis of the nucleon scalar form factor adopted to the case of a small
quark condensate. In this latter analysis, the shift of the scalar
form factor between the
Cheng-Dashen point and zero momentum transfer is found to be enhanced
over the result assuming strong quark condensation by up to a
factor of two, with substantial deviations starting to be visible for
$r=m_s/\hat{m}\lesssim 12$.\\[5mm]
PACS number(s):   12.39.Fe, 12.38.Lg, 11.55.Fv, 13.75.Gx

\newpage
\pagestyle{plain}
\newcommand{\gmu}{\mbox{$\gamma_\mu$}}
\newcommand{\gnu}{\mbox{$\gamma_\nu$}}
\newcommand{\psib}{\mbox{$\bar \Psi  $}}
\newcommand{\nablas}{\mbox{$\not \! \nabla  $}}
\newcommand{\chip}{\mbox{$\chi_{+} $}}
\newcommand{\chim}{\mbox{$\chi_{-} $}}
\newcommand{\bra}{\mbox{$\langle \, $}}
\newcommand{\ket}{\mbox{$\rangle  $}}
\newcommand{\nvb}{\mbox{$\bar  N_v $}}
\newcommand{\nv}{\mbox{$  N_v $}}
\newcommand{\hvb}{\mbox{$\bar H_v $}}
\newcommand{\hv}{\mbox{$  H_v $}}
\newcommand{\nbs}{\mbox{$ \not \! \nabla$}}
\newcommand{\vs}{\mbox{$\not \! v $}}
\newcommand{\nbts}{\mbox{$ \not\!\nabla^{\!\perp}$}}
\newcommand{\gf}{\mbox{$\gamma^5 $}}
\newcommand{\sig}{\mbox{$ \sigma^{\mu\nu}$}}
\newcommand{\eps}{\mbox{$\epsilon^{\mu\nu\rho\sigma} $}}
\newcommand{\us}{\mbox{$\not \! u $}}
\newcommand{\mh}{\mbox{$\hat m $}}
\newcommand{\fs}{\mbox{$F^2 $}}
\newcommand{\ga}{\mbox{$\st{\circ}{g}_A $}}
\newcommand{\gas}{\mbox{$\smash[t]{\overset{\circ}{g} }^2_A $}}
\newcommand{\mpi}{\mbox{$ M^{2}_{\pi}$}}
\newcommand{\bo}{\mbox{$B$}}
\newcommand{\lef}{\mbox{${\cal L}_{eff} $}}
\newcommand{\sigt}{\mbox{$\sigma (t)$}}
\newcommand{\dr}{\mbox{$\overrightarrow{D}^{\! n}_{\!\!\mu_1 \dots \mu_n}$}}
\newcommand{\drr}{\mbox{$\overrightarrow{D}^{\! n}_{\!\!\mu_n \dots
\mu_1}$}}
\newcommand{\dl}{\mbox{$\overleftarrow{D}^{\! n}_{\!\!\mu_1 \dots \mu_n}$}}
\newcommand{\dlr}{\mbox{$\overleftarrow{D}^{\! n}_{\!\!\mu_n \dots \mu_1}$}}
\newcommand{\dru}{\mbox{$\overrightarrow{D}^{\mu_1 \dots \mu_n}_{\! n}$}}
\newcommand{\Un}{\underline}
\newcommand{\ol}{\overline}
\newcommand{\ra}{\rightarrow}
\newcommand{\Ra}{\Rightarrow}
\newcommand{\ve}{\varepsilon}
\newcommand{\vp}{\varphi}
\newcommand{\vt}{\vartheta}
\newcommand{\dg}{\dagger}
\newcommand{\wt}{\widetilde}
\newcommand{\wh}{\widehat}
\newcommand{\br}{\breve}
\newcommand{\A}{{\cal A}}
\newcommand{\B}{{\cal B}}
\newcommand{\C}{{\cal C}}
\newcommand{\D}{{\cal D}}
\newcommand{\E}{{\cal E}}
\newcommand{\F}{{\cal F}}
\newcommand{\G}{{\cal G}}
\newcommand{\Ha}{{\cal H}}
\newcommand{\K}{{\cal K}}
\newcommand{\cL}{{\cal L}}
\newcommand{\M}{{\cal M}}
\newcommand{\N}{{\cal N}}
\newcommand{\cO}{{\cal O}}
\newcommand{\cP}{{\cal P}}
\newcommand{\R}{{\cal R}}
\newcommand{\cS}{{\cal S}}
\newcommand{\U}{{\cal U}}
\newcommand{\V}{{\cal V}}
\newcommand{\W}{{\cal W}}
\newcommand{\Y}{{\cal Y}}
\newcommand{\st}{\stackrel}
\newcommand{\del}{\partial}
\newcommand{\dint}{\displaystyle \int}
\newcommand{\dsum}{\displaystyle \sum}
\newcommand{\dprod}{\displaystyle \prod}
\newcommand{\dmax}{\displaystyle \max}
\newcommand{\dmin}{\displaystyle \min}
\newcommand{\dlim}{\displaystyle \lim}
\newcommand{\hy}{${\cal H}\! \! \! \! \circ $}
\newcommand{\h}[2]{#1\dotfill\ #2\\}
\newcommand{\tab}[3]{\parbox{2cm}{#1} #2 \dotfill\ #3\\}
\newcommand{\fsl}{\not\!}
\newcommand{\Fsl}{\not\!\!}
\def\lint{\int\limits}
\section{Introduction}

It is generally accepted that the chiral symmetry of massless QCD is
realized
in the Nambu-Goldstone mode. More precisely, it is ascertained that the
QCD vacuum spontaneously breaks chiral symmetry down to the diagonal
subgroup $U(3)_V$. This can actually been proven from first
principles if a vanishing $\theta$-vacuum is assumed, and provided
there are $N_f\geq 3$ massless flavours and
no coloured states in the spectrum (colour confinement).
\cite{tHooft80,VW84}
In accordance with Goldstone's theorem, eight
massless Goldstone bosons appear, each coupled via the coupling
constant $F_0$ to a conserved axial-vector current. The physics of these
Goldstone bosons can be described by a low energy effective theory,
called Chiral Perturbation Theory (ChPT) \cite{Weinberg79,GL84}. The masses
of the
 Goldstone
bosons are generated by explicit symmetry breaking terms in proportion
to $m_q$, the masses  of the light quarks $q=u,d$, and $s$.
Since $m_q$ is small compared to the typical mass scale
$\Lambda_H \approx 1$ GeV of the lightest massive hadrons not protected
by chiral symmetry, the effects of $m_q$ can be treated as a perturbation.

The coupling constant $F_0$ is an order parameter and a non-vanishing
value $F_0 \not= 0$ is a necessary and sufficient condition for
spontaneous breakdown of chiral symmetry (SB$\chi$S). The actual
mechanism of SB$\chi$S is not yet understood, however.
The light quark condensate in the chiral limit,
$< \bar{q} q >$, and the dimensionful parameter
\begin{equation}
B_0=-F^{-2} < \bar q q >
\end{equation}
play a special role in this respect. Two scenarios seem to be
theoretically viable: Large $B_0$ in the range of the mass scale
$\Lambda_H$, corresponding to strong condensation of quarks in the QCD
vacuum, or small $B_0$ in the range of $F_0$ (or even zero)
corresponding to SB$\chi$S realized via extended delocalized quarks with
high
``mobility''. \cite{Stern98} Although Lattice QCD simulations seem to
point towards a large quark condensate, there are other attempts
like in \cite{Kneur96} where a small condensate is reported. Given the
uncertainties inherent in these methods, it is fair to assume the
problem to be theoretically undecided. For a critical discussion of
the evidence resulting from QCD sum rules, we refer to the review article by
Stern \cite{Stern97}.

In the standard formulation of ChPT a large quark condensate, say
$B_0\approx\Lambda_H  $ is
assumed. The aim of generalized ChPT (GChPT) is to relax this
assumption, {\it i.e.} to allow  for small or even vanishing
$B_0$. \cite{stern91} Technically speaking, these assumptions amount to
different counting rules for the light quark masses and the  quark
condensate, {\it i.e.} \cite{stern91,stern94}
\begin{eqnarray}
m_q &=& {\cal O}(p^2), \quad B_0 = {\cal O}(1) \qquad\qquad {\rm
standard\ ChPT} \\
\label{Scounting}
m_q &=& {\cal O}(p), \ \quad B_0 = {\cal O}(p) \qquad\qquad {\rm
generalized\ ChPT},
\label{Gcounting}
\end{eqnarray}
where $p$ is a generic symbol for a low energy quantity. GChPT thus reorders
the
expansion of the effective lagrangian of low energy QCD. Summed up to
all orders it coincides with the standard approach. At each finite
order, however, GChPT takes into
account contributions which in the standard case are treated as higher
order terms. Since this reordering concerns the explicit symmetry
breaking sector only, the difference between
  standard and   generalized ChPT will be in proportion to the
light quark masses and hence small
\footnote{Recently, interesting consequences of a vanishing light
quark condensate have
been derived for the spectrum of vector- and axial-vector states in
the large $N_c$-limit of QCD. \cite{KdR97}}. The generalized approach allows
to
 experimentally  probe the size of the quark condensate \cite{stern94}.
The most promising case to
discriminate experimentally between the two scenarios is provided by
low-energy $\pi \pi$-scattering, \cite{pipi} where precise data should
be available in the near future. \cite{KLOEDIRAC}

Chiral symmetry also restricts the low energy interactions of pions
with baryons. Using the so called heavy baryon formalism (HBChPT),
the $\pi N$--system at low energies has been investigated extensively
and put to many tests. \cite{BKMrev} It is  natural to ask whether
it is possible to gain insight into the
mechanism of SB$\chi$S from the baryonic sector, as well. Incidentally,
the first evidence for a possible small light quark condensate was
obtained by an analysis of the Goldberger-Treiman discrepancy.
\cite{stern90}
However, this quantity turned out to be very sensitive to the
pion-nucleon coupling constant and therefore the analysis remained
inconclusive. Also, only the leading order corrections were
considered. Other single baryon processes like $\pi N \rightarrow N \pi$ or
$\gamma N \rightarrow N \pi \pi$, where abundant and precise data are
available, are potential candidates for testing the
assumption of large $B_0$ made in SChPT. In order to make such tests
possible,
however, HBChPT has to be adapted to the principles of the generalized
approach. Any calculation performed in SHBChPT can provide consistence
checks only, of course. The aim of the present article is to fill this
gap, {\it i.e.} to formulate generalized heavy baryon chiral
perturbation theory (GHBChPT). Two crucial assumptions will be
made. First, the quark mass counting rules of GChPT as given in Eq.
(\ref{Gcounting}) will be employed. This follows directly from the pure
Goldstone
Boson
sector. Second, we assume that the expectation values of non-singlet
operators between one-nucleon states scale with $\Lambda_\chi^D$,
where $\Lambda_\chi \approx 1$ GeV and $D$ denotes the canonical mass
dimensions of these correlation functions (no other small scales like
$B_0$ present). In particular, we treat dimensionless couplings like
$e_1$ in the effective $\pi N$--lagrangian
\begin{equation}
{\cal L}_{\pi N}=e_1 \bar\Psi {\rm tr} (\chi^\dagger U+U^\dagger \chi)
\Psi, \quad \chi=s+i p,
\label{pterm}
\end{equation}
as quantity of order unity. Here,
$\Psi$ denotes the nucleon field, $s$ and $p$ are scalar and
pseudoscalar sources, respectively, and $U$ contains the pion field in
the usual manner ({\it c.f.} section 2 for definitions).
The term in Eq. (\ref{pterm}) counts therefore as
order $p$.

Having made these assumptions, GHBChPT can be formulated along the
same lines as in the standard case. At each order in the chiral
expansion, the effective lagrangian contains additional terms compared
to the standard formulation. One of the main problems will be to obtain
estimates for these additional coupling constants. Given the many
observables available in the $\pi N$--system, the task is not hopeless.
It remains to be seen whether similar clean tests as those in $\pi
\pi$--scattering can be devised, ultimately leading to a better
understanding of the mechanism of spontaneous chiral symmetry
breakdown. This work is the first step of such a program.

 The article is organized as follows. In section 2 the effective
chiral lagrangian of GHBChPT is given to ${\cal O}(p^2)$ in it's most
general form and to ${\cal O}(p^3)$ in the scalar sector. Mass- and
wave-function renormalization to order $p^3$ are considered in
section 3. In section 4 we calculate the scalar form factor of the
nucleon to one-loop and compare with a dispersive theoretic  evaluation
adapted to the case of a small quark condensate.
Finally, we draw the conclusions in section 5. In Appendix A we
give a method to efficiently construct the relativistic baryonic effective
lagrangian for chiral SU(2) to all orders. Appendix B contains a
collection of loop functions employed in the article.

\section{The effective heavy baryon lagrangian  in the generalized approach}
\label{lagrange}

Our starting point is the QCD lagrangian with two massless quarks
coupled to external sources
\cite{GL84}
\footnote{In this paper we consider only chiral $SU(2)$. Accordingly, the
$SU(3)$ constants $F_0$ and $B_0$ are replaced by their $SU(2)$ counterparts
$F$ and $B$, respectively.}
\begin{equation}
\cL = \cL_{\rm QCD}^0 + \bar q \gamma^\mu \left(v_\mu + \frac{1}{3}
v_\mu^{(s)} + \gamma_5 a_\mu \right) q - \bar q (s - i \gamma_5 p)q,
\qquad q = \left( \begin{array}{c} u \\ d \end{array} \right)\; ,
\label{QCD}
\end{equation}
where $\cL_{\rm QCD}^0$ is the lagrangian of two-flavour QCD in the
absence of external sources.
The isotriplet vector (axial-vector) currents $v_{\mu}$ ($a_{\mu}$)
are hermitian and traceless matrix fields, whereas the current
$v_\mu^{(s)}$ is an isosinglet. The scalar and the pseudo-scalar
sources, $s$ and $p$ respectively, are also hermitian matrix fields in
isospin space, but in general are not traceless.
The lagrangian (\ref{QCD}) is symmetric
with respect to chiral transformations  $g\in G=SU(2)_L \times SU(2)_R$.
The QCD vacuum is assumed to spontaneously break chiral symmetry down
to the diagonal subgroup  $H=SU(2)_V $. According to Goldstone's
theorem, the spectrum of the theory contains three massless states
$\phi$, the pions, which are gathered in $u(\phi)$, an element of the
chiral coset space, {\it i.e.} $u(\phi) \in G/H$. Explicit symmetry
breaking terms in proportion to the light quark masses will give small
masses to the pions. Technically, this is incorporated by setting
the scalar source to $s = \M = \mbox{diag }(m_u,m_d)$ at the end of
the calculation.

The low-energy effective theory of pions and nucleons is obtained
by the CCWZ construction: chiral symmetry is realized
non-linearly on the Goldstone fields $\phi$ and matter fields $\Psi$
(nucleons) via \cite{CCWZ69}
\begin{align}
\label{trafo}
 u(\phi) &\,\st{g}{\ra} g_R u(\phi) h(g,\phi)^{-1} =
h(g,\phi) u(\phi) g_L^{-1} \, , \\
  \Psi = &\,{p \choose n} \st{g}{\ra} h(g,\phi) \Psi \, ,\\
  g =&\, (g_L,g_R) \in SU(2)_L \times SU(2)_R~. \nonumber
\end{align}
The compensator field $ h(g,\phi)\in SU(2)_V$ is defined by \eqref{trafo}
and
characterizes the non-linear realization. The effective lagrangian
consists of all invariants under chiral transformations which in
addition are hermitian and invariant under the discrete
transformations P, C, and T.
In order to construct these invariants explicitly, it is
convenient to define the following fields
\begin{equation}
\begin{split}
\label{ing}
u_\mu =&\, i \{ u^\dg(\partial_\mu - i r_\mu)u -
u(\partial_\mu - i \ell_\mu) u^\dg\}  \, ,
\\
\Gamma_\mu =&\, \frac{1}{2} \{ u^\dg (\partial_\mu - i r_\mu)u +
u (\partial_\mu - i \ell_\mu) u^\dg \} \, ,
\\
\chi_\pm =&\,  u^\dg \chi u^\dg \pm u \chi^\dg u ~,\qquad
\chi=   (s+ip)  \, ,
\\
f_\pm^{\mu\nu} =&\, u F_L^{\mu\nu} u^\dg \pm u^\dg F_R^{\mu\nu} u \, ,
 \\
v_{\mu\nu}^{(s)} =&\, \partial_\mu v_\nu^{(s)} - \partial_\nu v_\mu^{(s)}.
\\
\end{split}
\end{equation}
Here, the right and left handed fields $r_\mu = v_\mu + a_\mu$, $\ell_\mu =
v_\mu - a_\mu$
are the external gauge fields with associated non--Abelian field strengths
\begin{equation}
\begin{split}
F_R^{\mu\nu} =&\, \partial^\mu r^\nu - \partial^\nu r^\mu - i[r^\mu,r^\nu]
\, , \\
F_L^{\mu\nu}=&\, \partial^\mu \ell^\nu - \partial^\nu \ell^\mu -
i [\ell^\mu,\ell^\nu]~. \\
\end{split}
\end{equation}
Notice the missing factor $2B$ in the definition of $\chi_\pm$ in Eq.
(\ref{ing}). The  covariant derivative $\nabla_\mu$ acts on all fields
to the right of it and is formally given by
\begin{equation}\label{cov}
\begin{split}
\nabla_\mu &\,= \partial_\mu +
\left(\Gamma_\mu - i v_\mu^{(s)}\right)  \, ,
\\
\overleftarrow{\nabla}_{\!\!\mu} &\,= \overleftarrow{\partial}_{\!\!\mu} -
\left(\Gamma_\mu - i v_\mu^{(s)}\right) \, . \\
\end{split}
\end{equation}

We are now ready to discuss the consequences of generalized ChPT
for the pion-nucleon effective lagrangian. Before launching into the
discussion of generalized heavy baryon ChPT, however, let us recapitulate
some
facts about the Goldstone boson sector. The idea of a small or
vanishing quark condensate leads to a different counting rule for quark
mass terms, and consequently to a reordering of the chiral
lagrangian. Consider the expansion of the squared pion mass
\begin{equation}
\mpi= 2 B \hat m + 4 A \hat m^2 \, , \quad \hat m =\frac{1}{2}
\left( m_u+m_d \right) \; .
\label{Mpi2}
\end{equation}
  The constant $A$ is
an order parameter which has been estimated by chiral sum rules to
be of the order unity \cite{stern94}.
In the standard approach $B$ is assumed to be in the range set by the
hadronic mass scale $\Lambda_H$. Thus  the  first term
in (\ref{Mpi2}) is the leading contribution and the second term is
suppressed by a factor $m_q$.
 Equation \eqref{Mpi2}then implies the formal counting rule:
$B\sim {\cal O}(\Lambda_H)$ and $\mh \sim {\cal
O}(\frac{p^2}{\Lambda_H})$, where $p$ is a small
external momentum. Accordingly, the effective lagrangian is ordered in
local, chiral invariant terms ${\cal L}_{(k,l)}$, with $k$ powers of
covariant derivatives and $l$ powers of quark mass insertions \cite{stern94}
\begin{align}\label{xpt}
 \lef = &\,{\tilde { \cal L}}^{(2)} +{\tilde { \cal L}}^{(4)}+ \ldots \, ,
\\
\intertext{where }
{\tilde { \cal L}}^{(d)}= &\, \sum_{k+2l=d} { { \cal L}}_{(k,l)} \, .
\end{align}
The tilde on ${\tilde { \cal L}}^{(d)}$ signals that the associated coupling
constants are those of standard ChPT.

If $B$ is small, however, both terms in \eqref{Mpi2}  are
of the same order and, in general, equally important. The quark mass
$m_q$ and the coupling constant $B$, thus, must count as order ${\cal
O}(p)$.
The effective lagrangian $\lef$ must therefore be expanded not only in
powers of
covariant derivatives and quark mass insertions, but in powers of $B$ as
\mbox{ well \cite{stern91,stern94}}
\begin{align}\label{gxpt}
\lef = &\,{\cal L}^{(2)} +{\cal L}^{(3)}+ \ldots \, ,
\\
\intertext{where}
{\cal L}^{(d)}= &\, \sum_{j+k+l=d} B^j {\cal L}_{(k,l)} \, .
\end{align}
For instance, the leading order $p^2$ lagrangian of the Goldstone boson
sector consists of terms with either two covariant derivatives and no
mass insertions or of terms with no covariant derivatives and two mass
insertions
\cite{knecht97}
\begin{equation}
\begin{split}
{\cal L}^{(2)}_{\pi \pi} =
 \frac{F^2}{4}&\, \biggl[ \bra D_{\mu}U D^{\mu}U^{\dagger}\ket + 2B \,\bra
\chi^{\dagger}U+\chi U^{\dagger}\ket
\\
&\, A \,\bra \!\!\left( \chi^{\dagger}U\right)^2
 +\left( \chi U^{\dagger}\right)^2\ket
        +Z_p \,\bra \chi^{\dagger}U -U^{\dagger}\chi\ket^2
\\
&\, h_0\,\bra \chi^{\dagger}\chi\ket +h^{\prime}_{0} \left(
\det{\chi^{\dagger}}+\det{\chi}\right) \biggr] \, ,
\\
\end{split}
\label{Lpipi2}
\end{equation}
where  $U=u^2$ and the covariant derivative is defined as
\begin{equation}
        D_{\mu}U = \partial_{\mu} U -ir_{\mu}\, U
                                +iU \,l_{\mu} \, .
\end{equation}
Of course, Eq. (\ref{Lpipi2}) could be expressed in terms of the
chiral fields defined in (\ref{ing}).

To summarize, Eqs. \eqref{xpt} and \eqref{gxpt} state that summed to
all orders the standard and generalized approach coincide, namely
they describe the same effective lagrangian. To any finite order, however,
they differ, since in the generalized case terms are taken into
account which the standard case relegates to higher order.

Turning now to the baryonic sector,
our starting point is the generating functional for Green functions of
single nucleon processes defined by
\begin{equation}
e^{iZ[j,\eta,\bar\eta]} = {\cal N} \int [du d\Psi d \bar \Psi]
\exp [i\{ \tilde{S}_M + S_{MB} + \int d^4 x (\bar \eta \Psi
+ \bar \Psi \eta)\}]~.
\label{Z}
\end{equation}
$\tilde{S}_M$ and $S_{MB}$ are the mesonic and relativistic
pion-nucleon effective actions, respectively. The tilde on $\tilde{S}_M$
accounts for the fact that in (\ref{Z}) the nucleon degrees of
freedom have not yet been integrated out. The form of the mesonic
action remains the same, {\it c.f.} the leading order expression given
in (\ref{Lpipi2}) above.  The coupling constants in general are
different, however, since they might get contributions from closed
nucleon loops. \cite{EM97} In the pion-nucleon sector the relativistic
lowest-order chiral lagrangian takes the form
\cite{GSS88}
\begin{equation}
\begin{split}
\cL_{\pi N}^{(1)} = \bar \Psi \biggl[
&\,i \not\!\nabla - m + \frac{\ga}{2}\not\!u \gamma_5
+  e_1 \bra \chip\ket
+ e_2\bigl( \chip -\frac{1}{2}\bra \chip\ket\bigr)
\\
&\, + e_3 \gamma_5 \bra \chim\ket
  + e_4 \gamma_5   \chim
 \biggr]\Psi\, . \label{LMB1}
\\
\end{split}
\end{equation}
A characteristic feature of generalized ChPT is the appearance of
terms in proportion to $e_1, \ldots e_4$ already at leading order.
In standard ChPT they are present only at next-to-leading order.
We have chosen new symbols for the corresponding
coupling constants in order not to confuse these couplings with those
of standard ChPT.

Next we take the non-relativistic limit in order to get rid of the large
nucleon mass term, thereby allowing for a consistent low energy expansion.
\cite{MRR92,BKKM92,Eck94}
Projecting the nucleon field $\Psi$ onto its velocity dependent
light  and heavy components  $\nv$ and $\hv$ respectively \cite{Ge90}
\begin{equation}
\begin{split}
\label{vdf}
N_v(x) =&\, \exp[i m v \cdot x] P_v^+ \Psi(x) \, , \\
H_v(x) =&\, \exp[i m v \cdot x] P_v^- \Psi(x) \, ,  \\
P_v^\pm =&\, \frac{1}{2} (1 \pm \not\!v)~, \qquad v^2 = 1 \; , \\
\end{split}
\end{equation}
the pion--nucleon Lagrangian   is brought to the form
\begin{equation}\label{piN}
\cL_{\pi N} = \bar N_v A N_v + \bar H_v B N_v + \bar N_v \gamma^0 B^\dg
\gamma^0 H_v - \bar H_v C H_v \, .
\end{equation}
The  mesonic field operators $A$, $B$, $C$ admit chiral expansions of
the form
\begin{align}
\label{A1}
\begin{split}
A =&\, A_{(1)} + A_{(2)} + A_{(3)} + \ldots  \, ,  \quad
B =  B_{(1)} + B_{(2)}   + \ldots  \, , \\
\qquad
&\,\qquad \qquad\qquad C =  C_{(0)} + C_{(1)}   + \ldots  \, , \\
\end{split}
\end{align}
where $A_{(n)}$ denotes a quantity of order $p^n$. Explicitly,
the leading order expressions are given by
\begin{align}
\begin{split}
A_{(1)} =&\, iv \cdot \nabla + \ga S \cdot u + e_1\bra \chip \ket
 + e_2\bigl( \chip -\frac{1}{2}\bra \chip\ket\bigr)\, \, ,
\\
B_{(1)}=&\, i \not\!\nabla^\perp
 - \dfrac{\st{\circ}{g}_A}{2} v \cdot u \gamma_5
 +  e_3 \gamma_5 \,\bra \chim\ket
 + e_4 \gamma_5   \chim   \, ,
\\
C_{(0)} =&\, 2m \, ,\\
 C_{(1)}=&\,  i v \cdot \nabla + \st{\circ}{g}_A S \cdot u
- e_1\bra \chip \ket
- e_2\bigl( \chip -\dfrac{1}{2}\bra \chip\ket\bigr) \, ,
\\
 \nabla_\mu^\perp=&\, \nabla_\mu - v_\mu v \cdot \nabla \, , \qquad
S^\mu = i \gamma_5 \sigma^{\mu\nu} v_\nu/2 \, .
\\
\end{split}
\end{align}
Performing a Gaussian integration in the generating functional,
the fields $H_v$ are integrated out. Introducing
sources corresponding to $N_v$ and $H_v$, respectively,
\begin{equation}
\rho_v=e^{i m v\cdot x} P_v^+ \eta, \qquad R_v=e^{i m v\cdot x} P_v^-
\eta \; ,
\end{equation}
and shifting also the variable $N_v$, the generating functional
can be brought to the form
\begin{eqnarray}
e^{iZ[j,\eta,\bar\eta]} &=& {\cal N} \int [du d N_v d \bar N_v]
\exp \biggl[i \tilde{S}_M \nonumber \\
& & + i \int d^4x\;
\Bigl\{ \bar N_v (A + \gamma^0 B^\dg \gamma^0 C^{-1} B) N_v +
\bar R C^{-1} R  \nonumber\\
& &
\ - (\bar\rho+\bar R C^{-1} B) (A + \gamma^0 B^\dg \gamma^0 C^{-1} B)
(\rho+\gamma^0 B^\dg \gamma^0 C^{-1} R) \Bigr\} \biggr] \; .
\label{Zintermed}
\end{eqnarray}
In HBChPT, the matrix $C^{-1}$ in (\ref{Zintermed}) is expanded in a
power series in $1/m$ before performing the functional integral over
$N_v$. Then, this functional integral yields only a constant. However,
as pointed out in \cite{EM97}, we know the effect of the interchange
of limits must be to change $\tilde{S}_M$ into $S_M$, with $S_M$ the usual
effective action of ChPT. The standard
procedure of ChPT can be applied, \cite{GL84} and we refer to ref.
\cite{EM97} for the relation of the T-matrix elements obtained in
HBChPT and the fully relativistic S-matrix elements one is actually
looking for. We come back to this point in section 3 when we consider
mass and wavefunction renormalization.

Here we are concerned only with the effective low energy action of the
pion-nucleon system resulting from the analysis above
\begin{equation}
S_{\pi N} = \int d^4x \; \wh\cL_{\pi N} = \int d^4x \; \bar N_v
(A + \gamma^0 B^\dg \gamma^0 C^{-1} B) N_v~,
\end{equation}
where $C^{-1}$ is understood to be expanded in inverse powers of
the nucleon mass. We thus obtain
the chiral Lagrangian of GHBCHPT up to ${\cal O}(p^3)$ in its general
form
\begin{equation}
\begin{split}
\label{ABC}
 A +\,\gamma^0 B^\dg \gamma^0 C^{-1} B &= A_{(1)}
\\
\phantom{A +\,\gamma^0 B^\dg \gamma^0 C^{-1} B }&+ A_{(2)} +
\frac{1}{2m} \gamma^0 B^\dg_{(1)} \gamma^0 B_{(1)}
\\
\phantom{A +\,\gamma^0 B^\dg \gamma^0 C^{-1} B }&
+ A_{(3)} + \frac{1}{2m} \left(\gamma^0 B^\dg_{(2)} \gamma^0
B_{(1)} + \gamma^0 B^\dg_{(1)} \gamma^0 B_{(2)}\right)
\\
\phantom{A +\,\gamma^0 B^\dg \gamma^0 C^{-1} B }&- \frac{1}{4m^2}
\gamma^0 B^\dg_{(1)} \gamma^0 C_{(1)} B_{(1)} + {\cal O}(p^4)\, .
\\
\end{split}
\end{equation}

In the following we detail the derivation of the explicit effective
pion-nucleon lagrangians up to order $p^3$. A recipe of how to
construct the relativistic invariants for chiral SU(2) to all orders
in the chiral expansion is given in App. A. We follow closely the
presentation given in ref. \cite{EM96}.

At lowest order the lagrangian is given by $A_{(1)}$
\begin{equation}
\wh \cL_{\pi N}^{(1)} =  \bar N_v\left[iv \cdot \nabla + \ga S \cdot u +
                         e_1\bra \chip \ket+
                           e_2\left( \chip -\frac{1}{2}\bra \chip\ket\right)
                             \right] N_v~ \, .
\label{piN1}
\end{equation}
The operators associated with the coupling constants $ e_1$ and $ e_2$ are a
new feature of the generalized approach. In the standard case these two
operators
are  treated as operators of the order ${\cal O}(p^2)$ and they are
associated with the couplings $a_3$ and $a_4$ respectively \cite{EM96}.

The relativistic chiral lagrangian at order ${\cal O}(p^2)$ has four types
of contributions:
\begin{equation}\label{lag2}
   \cL_{\pi N}^{(2)}=  \cL_{(2 , 0)}+  \cL_{( 1, 1)}+
\cL_{( 0, 2)} +B \,  \cL_{\pi N}^{(1)} \, .
\end{equation}
The  term $  \cL_{(2 , 0)}$ is the same as in the standard case. The
appearance of  $  \cL_{( 1, 1)}$ and of $  \cL_{( 0, 2)}$ already at the
order ${\cal O}(p^2)$ is  characteristic of the generalized approach.
 $  \cL_{( 1, 1)}$ can be read off from the ${\cal O}(p^3)$ lagrangian
given in \cite{EM96} and $  \cL_{( 0, 2)}$ counted formally as an
${\cal O}(p^4)$
contribution.
 Since the loop divergences associated with virtual pion and nucleon
exchange
  are at least of order ${\cal O}(p^3)$,   the
last term in \eqref{lag2} only leads to a finite renormalization of
 $ \cL_{\pi N}^{(1)}$ and may be neglected.
In order to construct the effective lagrangian at ${\cal O}(p^2)$ we have
used
the following set of operators \cite{EM96}:
\begin{eqnarray}
\label{P2}
\langle u_\mu u^\mu\rangle,  &\ra& A_{(2)} \nonumber
\\
i \sigma^{\mu\nu} u_\mu u_\nu, \;
\sigma^{\mu\nu} f_{+\mu\nu}, \;
\sigma^{\mu\nu} v_{\mu\nu}^{(s)}, \;
\langle u_\mu u_\nu \rangle
i \gamma^\mu \nabla^\nu + {\rm h.c.}
&\ra& A_{(2)}, B_{(2)} \nonumber
\\
\left[\not\!u \, , \chim\right] ,\;
\gamma_5 \not\!u \bra \chip \ket ,\;
\gamma_5 \gamma_{\mu} \bra u^{\mu} \chip \ket ,\;
\gamma_5 \left[\not\!\nabla ,\chim\right],\;
\gamma_5 \gamma_{\mu} \bra \partial^{  \mu} \chim \ket
&\ra& A_{(2)} ,B_{(2)}
\\
\bra \chip \ket^2 ,\; \chip \bra \chip \ket ,\; \bra \chip \chip \ket ,\;
\bra \chim \ket^2 ,\; \chim \bra \chim \ket ,\; \bra \chim \chim \ket \;
&\ra& A_{(2)} \, .\nonumber
\end{eqnarray}
In the derivation of  \eqref{P2} we have omitted all operators, which
can be eliminated with a redefinition of the nucleon field $\Psi$ as
given in the appendix \ref{appendixa},
{\it e.g.} the term
 $$ \langle u_\mu u_\nu\rangle \nabla^\mu \nabla^\nu + {\rm h.c.}$$ can be
eliminated with equation \eqref{two}.
The two pieces at order ${\cal O}(p^2)$ in  \eqref{ABC} are then found to be
\begin{align}
\begin{split}\label{a2}
A_{(2)} =& \,  \frac{c_2}{m}
(v \cdot u)^2 + \frac{c_3}{m} u \cdot u +
   \frac{1}{m} \ve^{\mu\nu\rho\sigma} v_\rho S_\sigma
\left[i c_4 u_\mu u_\nu + c_6 f_{+\mu\nu} +c_7 v^{(s)}_{\mu\nu}\right]
\\
&\,+\frac{c_8}{m} [\chi_-, v \cdot u]
+\frac{c_9}{m}i S^\mu [\nabla_{\!\mu},\chi_-]
+\frac{c_{10 }}{m}i S^\mu \langle \partial_{\mu} \chi_-\rangle
\\
&\,
+\frac{c_{11}}{m}S \cdot u \langle \chi_+\rangle
+\frac{c_{12}}{m}S^\mu \langle u_\mu \chi_+ \rangle
\\
&\,+\frac{c_{13 }}{m}\bra \chip \ket^2
+ \frac{c_{14 }}{m}\chip \bra \chip \ket
+ \frac{c_{15 }}{m}\bra \chip \chip \ket \\
&\,
+\frac{c_{16 }}{m} \bra \chim \ket^2
+ \frac{c_{17 }}{m}\chim \bra \chim \ket
+ \frac{c_{18 }}{m}\bra \chim \chim \ket  \, ,
\end{split}
\end{align}
and
\begin{align}
\begin{split}\label{B1B1}
 \frac{1}{2m} \gamma^0 B^\dg_{(1)} \gamma^0 B_{(1)} = &\,\frac{1}{2m}
 \biggl[ (v \cdot \nabla)^2 - \nabla \cdot \nabla - i\ga \{S \cdot \nabla,
v \cdot u\}
 \\
&\, \mbox{}\quad - \frac{\gas}{4} (v \cdot u)^2 + \frac{1}{2}
\ve^{\mu\nu\rho\sigma} v_\rho S_\sigma \left[i u_\mu u_\nu + f_{+\mu\nu}
+ 2 v^{(s)}_{\mu\nu}\right]
\\
&\,\quad-\frac{1}{2}\ga  e_4[\chi_-, v \cdot u]
+2e_4i S^\mu [\nabla_{\!\mu},\chi_-]
+2e_3i S^\mu \langle \partial_{\mu} \chi_-\rangle
\\
&\,\quad+ e^2_3 \bra \chim \ket^2
+ e^2_4  \chim \chim
+2 e_3  e_4\chim \bra \chim \ket \biggr]  \, .
\end{split}
\end{align}
The operator $(v\nabla)^2/2m$ in \eqref{B1B1} is a special case of an
equation-of-motion type term. Up to ${\cal O}(p^3)$ we have encountered
the following equation-of-motion type terms  \cite{EM96}:
 \begin{equation}
\cL_{\text{EOM}} =
\bar N_v \left\{ X\left(iv \cdot \nabla\right)^3 + iv \cdot
\overleftarrow{\nabla}
Y iv \cdot \nabla + Z iv \cdot \nabla - iv \cdot \overleftarrow{\nabla}
\gamma^0 Z^\dg \gamma^0 \right\}
N_v~  \, .\label{EOM}
\end{equation}
The operators $Y=Y^{\dagger}$ and $Z$ are purely mesonic operators and are
at most of the order ${\cal O}(p)$ and ${\cal O}(p^2)$ respectively and $X$
is a real constant.
Applying an
adaptation of the field redefinition given in \cite{EM96}
\begin{equation}
\begin{split}
N_v =&\, \biggl\{ 1 - \frac{X}{2} (iv \cdot \nabla)^2 + \frac{1}{2}
\left[ Y +\Delta {\cal L}^{(1)}\right]iv \cdot \nabla
+ \frac{X}{2} \left[iv \cdot \nabla, \Delta {\cal L}^{(1)}\right]
\\
     &\, \mbox{} - \frac{X}{2} \left(\Delta {\cal L}^{(1)} \right)^2
-\frac{Y}{2}\Delta {\cal L}^{(1)}
 - \gamma^0 Z^\dg \gamma^0 \biggr\} N'_v \,
\label{nft}
\end{split}
\end{equation}
with
\begin{equation*}
\Delta {\cal L}^{(1)} = \ga  S \cdot u  + e_1\bra \chip \ket
                   + e_2\left( \chip -\frac{1}{2}\bra \chip\ket \right)
\end{equation*}
to the leading order lagrangian $\wh \cL_{\pi N}^{(1)}$ eliminates all
equation-of-motion terms. It also induces
a lagrangian at order ${\cal O}(p^2)$ which in our case with $X=Z=0$ and
$Y=1/(2m)$ is explicitly given by
\begin{equation}
{\cal L}_{\text{ind}}=-\frac{1}{2m}\bar{N_v'} (\Delta {\cal L}^{(1)})^2
N_v'\, .
\end{equation}
Naturally this field redefinition induces a lagrangian at order ${\cal
O}(p^3)$ which must be dealt with in the construction of $\wh \cL_{\pi
N}^{(3)}$.
Adding all the pieces  together
 the effective lagrangian at order ${\cal O}(p^2)$ is then given in its
final form by
\begin{equation} \label{LpiN2}
\begin{split}
\wh \cL_{\pi N}^{(2)} = \bar N_v&\,\biggl[ - \frac{1}{2m} \left(\nabla \cdot
\nabla + i\ga \{S \cdot \nabla, v \cdot u\}\Big) \right.
 \\
&\, \mbox{} + \frac{ f_2}{m} \langle (v \cdot u)^2\rangle
 +\frac{ f_3}{m} \langle u \cdot u\rangle
 \\
&\,  \mbox{} + \frac{1}{m} \ve^{\mu\nu\rho\sigma} v_\rho S_\sigma
[i  f_4 u_\mu u_\nu
+  f_6 f_{+\mu\nu}
+  f_7 v_{\mu\nu}^{(s)}]
\\
&\,  +\frac{ f_{8}}{m} \left[ \chi_-, v \cdot u\right]
 + \frac{ f_{9}}{m}iS^{\mu} \left[\nabla_{\!\mu},\chim \right]
+\frac{ f_{10}}{m}iS^{\mu} \bra\partial_{\mu}\chim \ket
\\
&\,  +\frac{ f_{11}}{m} S\cdot u \bra\chip\ket
+\frac{ f_{12}}{m} S^{\mu} \bra u_{\mu}\chip\ket
\\
&\,+\frac{ f_{13}}{m}\bra \chip\ket^2
+ \frac{ f_{14}}{m}\chip \bra \chip\ket+
\frac{ f_{15}}{m}\bra \chip^2 \ket
\\
&\,+\frac{ f_{16}}{m}\bra \chim\ket^2
+ \frac{ f_{17}}{m}\chim \bra \chim\ket+
\frac{ f_{18}}{m}\bra \chim^2 \ket
\biggr] N_v~ \, .
\\
\end{split}
\end{equation}
The coupling constants $f_i$ are related to those appearing previously by
\begin{alignat*}{3}
f_2 &= \frac{c_2}{2}-\frac{\gas}{8} &\qquad
f_3 &=  \frac{c_3}{2}+\frac{\gas}{16}&\qquad
f_4 &= c_4+\frac{1-\gas}{4}
\\
f_6& = c_6 +\frac{1}{4} &\qquad
f_7& = c_7 +\frac{1}{2} &\qquad
f_8 &= c_8 - \frac{1}{4}\ga e_4
\\
f_9 &= c_9 +e_4  & \qquad
f_{10} &= c_{10} + e_3  & \qquad
f_{11} &= c_{11} -\ga e_1
\\
f_{12} & = c_{12}-\frac{\ga e_2}{2}& \qquad
f_{13} & =c_{13}-\frac{e^2_1}{2} +\frac{e_1e_2}{2} +\frac{e^2_2}{8}& \qquad
f_{14} & =c_{14}-e_1e_2
\\
f_{15} &= c_{15}-\frac{e^2_4}{4}& \qquad
f_{16} & = c_{16} +\frac{e^2_3}{2} -\frac{e^2_4}{4} & \qquad
f_{17} & = c_{17} + e_3 e_4 +\frac{e^2_4}{2}
\\
&&\qquad f_{18} & = c_{18} +\frac{e^2_4}{4} \, .&\qquad &
\end{alignat*}

At order ${\cal O}(p^3)$ we restrict ourselves   to   operators
which contribute to either mass- and wave function renormalization
of the nucleon field $\Psi$ or to the scalar sector of the $\pi
N$-system. This means that we set all chiral fields to zero except for
the scalar source $\chip$. The non relativistic lagrangian
$\wh \cL_{\pi N}^{(3)}$ has the following contributions:
\begin{enumerate}
\item
As stated before the generalized approach is also an expansion in the
parameter $B$. Therefore the relativistic lagrangian at order ${\cal
O}(p^3)$ includes the two terms of the form
\begin{equation}
B^2 \, \cL_{\pi N}^{(1)} +B \, \cL_{\pi N}^{(2)} \, .
\end{equation}
In the non-relativistic limit their $B$-dependent contributions have
the same chiral structure as $A_{(1)}$ and $A_{(2)}$, but have new
coupling constants.
These $B$-dependent counterterms of ${\cal O}(p^3)$ are needed to
renormalize divergences that arise from using the vertices of
$\wh \cL_{\pi N}^{(1)}$ in the loop. They renormalize the coupling
constants of $\wh \cL_{\pi N}^{(1)}$ by contributions of ${\cal
O}(B^2)$ and the  constants of $\wh \cL_{\pi N}^{(2)}$
by an amount of  ${\cal O}(B)$. As can bee seen from the lowest order
lagrangian,
$B$ enters only via the product $\mh B$, thus there is no loop-divergence
in proportion to $B^2 \mh$. We can therefore omit all operators of the form
$B^2 \, \cL_{\pi N}^{(1)}$.
\item The relativist lagrangian $ \cL_{\pi N}^{(3)}$ contains genuine new
operators that contribute in the non-relativistic limit to $A_{(3)}$, see
equation \eqref{ABC}:
\begin{equation}\label{a3}
\begin{split}
\sum^{7}_{i=1}{\tilde g}_i {\cal O}_i=
&\,\frac{{\tilde g}_1}{m^2} \left[\nabla_{\!\mu},\left[ \nabla^{\mu}, \chip
\right]\right]
+\frac{{\tilde g}_2}{m^2}\bra \partial \!\cdot \!\partial\,\chip \ket
+\frac{{\tilde g}_3}{m^2}\chip  \bra \chip\chip \ket
\\
+&\,
\frac{{\tilde g}_4}{m^2}\chip  \bra \chip \ket^2
+\frac{{\tilde g}_5}{m^2}\bra \chip \ket^3
+\frac{{\tilde g}_6}{m^2}\bra \chip \chip\ket \bra \chip \ket \, .
\\
+&\,
\frac{{\tilde g}_7 }{m^2}
\left[\chip,\left[iv\cdot \nabla,\chip \right]\right]\, .
\end{split}
\end{equation}
  Taking into account the terms in proportion to $B$ we find for  $A_{(3)}$
\begin{equation}
\begin{split}
A_{(3)}=&\, \sum^{7}_{i=1}{\tilde g}_i {\cal O}_i+
\frac{B}{m^2}\left[
\tilde{c}_{13 }\bra \chip \ket^2
+ \tilde{c}_{14 }\chip \bra \chip \ket
+ \tilde{c}_{15 }\bra \chip \chip \ket
\right] \, .
\end{split}
\end{equation}
\item
The contributions from the $1/m$ expansion in \eqref{ABC}. The terms
proportional to $B_{(2)}$, however, do not contribute in our case.
\item  The field redefinition as given in \eqref{nft} with $X=Z=0$
and $Y=1/(2m)$ applied to
$\wh \cL_{\pi N}^{(1)}$  and to the sum of the equations \eqref{a2}
and \eqref{B1B1} induces a lagrangian at
${\cal O}(p^3)$, which must be taken into account. After this
transformation there are still equation-of-motion terms at the
${\cal O}(p^3)$ level. These can be removed by a second transformation
inducing additional terms to the
${\cal O}(p^3)$-lagrangian. The appropriate choice of $X,Y$ and $Z$
will be given below.
\end{enumerate}
In a first step we collect all contributions from $1\dots 3$ and from the
field transformation \eqref{nft}. Splitting the  ${\cal O}(p^3)$-lagrangian
into
equation-of-motion terms and a remainder we find:
\begin{equation}
\wh \cL_{\pi N}^{(3)}= \wh \cL_{\text{EOM}}^{(3)}+\wh \cL_{\text{rem}}^{(3)}
\end{equation}
 with
\begin{equation} \label{eqm}
\wh \cL_{\text{EOM}}^{(3)}= \frac{1}{16m^2} {\bar N}_v'
\left\{ \bar X \left(iv \cdot \nabla\right)^3 + iv \cdot
\overleftarrow{\nabla}
\bar Y iv \cdot \nabla + \bar Z iv \cdot \nabla - iv \cdot
\overleftarrow{\nabla} \gamma^0 {\bar Z}^\dg \gamma^0 \right\}
N_v' \, ,
\end{equation}
 where
\begin{equation}
\begin{split}\label{xyz}
\bar X = &\, 1\, ,
\\
\bar Y = &\, {\Delta \cal L}^{(1)} \, ,
\\
\bar Z = &\,  \frac{1}{2}\left({\Delta \cal L}^{(1)}\right)^2
+  4{\Delta \cal L}^{(2)}
-\left[ iv \cdot \overrightarrow{\nabla},{\Delta \cal L}^{(1)}\right]
\, ,
\end{split}
\end{equation}
and
\begin{equation}\label{noneqm}
\begin{split}
\wh \cL_{\text{rem}}^{(3)}= \frac{1}{16m^2}{\bar N}_v'\biggl\{
&\,16m^2\sum^{7}_{i=1}{\tilde g}_i {\cal O}_i
+\frac{1}{2} \left[ \Delta \cL^{(1)} , \left[ i v\cdot \nabla , \Delta
\cL^{(1)}\right]\right]
+\left(  \Delta \cL^{(1)}\right)^3
\\-&\,
4 \left\{ \Delta \cL^{(1)},\Delta \cL^{(2)}\right\}
+2\left[ \nabla_{\!\mu},\left[\nabla^{\mu}, \Delta \cL^{(1)}\right]\right]
\\+&\,
4 i \epsilon^{\mu\nu\rho\sigma}v_{\rho}S_{\sigma}\biggl[
\overleftarrow{\nabla}_{\!\!\mu} \left[ \nabla_{\!\!\nu}\, ,\Delta
\cL^{(1)}\right]
-\left[ \nabla_{\!\!\nu}\, ,\Delta \cL^{(1)}\right]\nabla_{\!\mu}\biggr]
\\
+&\,
16B \left[
\tilde{c}_{13 }\bra \chip \ket^2
+ \tilde{c}_{14 }\chip \bra \chip \ket
+  \tilde{c}_{15 }\bra \chip \chip \ket \right]
\biggr\} N_v' \, ,
\end{split}
\end{equation}
 where in our case ${\Delta \cal L}^{(1)}$ reduces to
\begin{equation}
\Delta \cL^{(1)}=  e_1 \bra \chip \ket +e_2\left( \chip -\frac{1}{2}
\bra \chip \ket \right) \, ,
\end{equation}
 and  ${\Delta \cal L}^{(2)}$ is given by
\begin{equation}
{\Delta \cal L}^{(2)}=
  c_{13} \bra \chip\ket^2+c_{14} \chip \bra \chip\ket
+c_{15} \bra \chip^2\ket \, .
\end{equation}
The terms in Eqs. \eqref{xyz} and
\eqref{noneqm} receive contributions from
 the transformation \eqref{nft} when applied
  to $\wh \cL_{\pi N}^{(1)}$ and to  the sum of the equations
   \eqref{a2} and \eqref{B1B1}. The  $1/m$ expansion contributes as well.

In order to eliminate $\wh \cL_{\text{EOM}}^{(3)}$ we  employ \eqref{nft}
with
 $$X=\frac{1}{16m^2}\, \bar X\, ,\quad Y=\frac{1}{16m^2}\,\bar Y \,
\quad\text{and} \,\quad Z=\frac{1}{16m^2}\bar Z\, .$$
The induced lagrangian is readily obtained:
\begin{equation}
\cL_{\text{ind}}= \frac{1}{16m^2}\bar N_v' \biggl[
 -(\Delta \cL^{(1)})^3
-\frac{1}{2} \left[ \Delta \cL^{(1)} , \left[ i v\cdot \nabla , \Delta
\cL^{(1)}\right]\right]
 -4 \left\{ \Delta \cL^{(1)},\Delta \cL^{(2)}\right\}
\biggr] N_v'\, .
\end{equation}
Adding everything together we find the lagrangian in the scalar
sector to ${\cal O}(p^3)$
\begin{equation}
\begin{split} \label{LpiN3}
\wh \cL_{\pi N}^{(3)}= \nvb \Biggl\{
&\,\frac{g_1}{m^2} \left[\nabla_{\!\mu},\left[ \nabla^{\mu},
\chip\right]\right]
+\frac{{g}_2}{m^2}\bra \partial \!\cdot \!\partial\,\chip \ket
+\frac{{g}_3}{m^2}\chip  \bra \chip\chip \ket
+\frac{{g}_4}{m^2}\chip  \bra \chip \ket^2
\\
+&\,
\frac{{g}_5}{m^2}\bra \chip \ket^3
+\frac{{g}_6}{m^2}\bra \chip \chip\ket \bra \chip \ket \,
+\frac{{g}_7 }{m^2} \left[\chip,\left[iv\cdot \nabla, \chip \right]\right]
\\
+&\,
\frac{1}{4m^2}\,i \epsilon^{\mu\nu\rho\sigma}v_{\rho}S_{\sigma}\Biggl[
e_1\biggl(\overleftarrow{\nabla}_{\!\!\mu} \bra\partial_{\nu}\chip\ \ket
-\bra\partial_{\nu}\chip\ \ket\nabla_{\!\mu}\biggr)
\\+&\,
e_2\biggl(\overleftarrow{\nabla}_{\!\!\mu}
       \bigl[\nabla_{\!\nu}\,,\chip-\dfrac{1}{2}\bra\chip \ket\bigr]
      -\bigl[\nabla_{\!\nu}\,,\chip-\dfrac{1}{2}\bra\chip \ket\bigr]
\nabla_{\!\mu}\Biggr]
\\
+&\,
\frac{B}{m^2}\left[
\tilde{c}_{13 }\bra \chip \ket^2
+ \tilde{c}_{14 }\chip \bra \chip \ket
+  \tilde{c}_{15 }\bra \chip \chip \ket \right]
\Biggr\} N_v  \, .
\end{split}
\end{equation}
In \eqref{LpiN3} we have subsumed all operators of the form given in
\eqref{a3}
into the coupling constants $g_i$. They differ from  ${\tilde g}_i$ by a
finite
renormalization.

\section{Mass-- and wavefunction renormalization}

The formalism presented in section 2 enables us to calculate the
T-matrix element of any process with one incoming and one outgoing
nucleon in Generalized HBChPT to order $p^3$. However, as shown by
Ecker and Moj\v zi\v s, the sources of the heavy
component of the nucleon field in the generating functional cannot
be dropped altogether. \cite{EM97} Rather, these terms contribute to
the wavefunction renormalization in a non-trivial manner. In order to
provide this link between the T-matrix elements calculated in GHBChPT
and the relativistic S-matrix elements one is actually seeking, we
discuss mass-- and wavefunction renormalization to order $p^3$. The
formalism of ref. \cite{EM97} can be carried over directly to our
case and need not to be repeated here. However, we indicate
those steps where Generalized ChPT gives raise to new features
not present in the standard formulation. In the following we work in
the isospin limit.

\subsection{The nucleon propagator in generalized ChPT}

The central object to be considered is the nucleon propagator ({\it
c.f.} ref. \cite{EM97} for definitions)
\begin{equation}
S_N(p)=P_v^+ S_{++} P_v^+ + P_v^+ S_{+-} P_v^- +
       P_v^- S_{-+} P_v^+ + P_v^- S_{--} P_v^-
\label{SN}
\end{equation}
where the off--shell momentum $p$ is decomposed according to
\begin{equation}
p=m v+k
\end{equation}
with $k$ a residual small momentum. For the present application, we
need the pole-part of the objects $S_{ij}$, $i,j \in \{+,-\}$.

\begin{figure}[p]
\begin{center}
\leavevmode
\epsfig{file=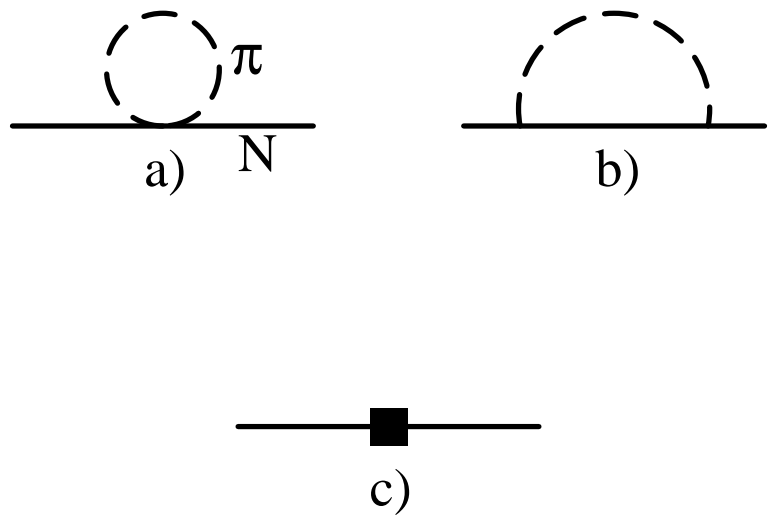,width=10cm}
\caption[]{
 Feynman diagrams contributing to the  self-energy of the nucleon to
${\cal O}(p^3)$. Plain and  dashed lines  denote   the nucleon and the
pion, respectively. The  shaded box  denotes the counterterm insertions
of order $p^2$ and $p^3$.} \label{self}
\end{center}
\end{figure}

$S_{++}$ in (\ref{SN}) is determined by the selfenergy calculated in
GHBChPT
\begin{equation}
S_{++}(k)^{-1}=- \Sigma(k) \; .
\end{equation}
The diagrams which contribute are shown in Fig. \ref{self}. Diagram
a) is a typical new feature of the generalized framework --- in
standard HBChPT these diagrams enter first at order $p^4$.
Diagram b) is formally the same as in the standard case, but
there are nevertheless some differences. First, the propagator in
GHBChPT is modified to
\begin{equation}
S^{\rm GHBChPT}= {i \over v\cdot k-\sigma_0},
\qquad \sigma_0\equiv -4 e_1 \hat m \; .
\label{HBprop}
\end{equation}
We will later see that the shift $\sigma_0$ in the propagator gives raise
to higher order contributions only and has no net effect at order
$p^3$.
Second, the pion mass entering via the pion propagator is given by the
leading
order expression in Eq. \eqref{Mpi2}.
Here and below, the symbol $M_\pi^2$ is always understood to be defined by
that equation.
Finally, the contact terms of diagram c) can be read off from
Eqs. (\ref{LpiN2}) and (\ref{LpiN3}). Explicitly, the selfenergy is
found to be
\begin{equation}
\Sigma(k)= \Sigma_{\rm loop}(v\cdot k-\sigma_0)+\Sigma_{\rm contact}
\end{equation}
with
\begin{equation}
\Sigma_{\rm loop}(\omega)=\Sigma^{(a)}+\Sigma^{(b)}(\omega)
\label{Sigmaloop}
\end{equation}
and
\begin{eqnarray}
\Sigma^{(a)}&=&-{3 \over 2} {\sigma_0 \over F^2} \frac{1}{i}\Delta
(M_\pi^2)\\
\Sigma^{(b)}(\omega)&=&{3 \over 4} {g_A^2\over F^2} \left(
-\omega \frac{1}{i}\Delta (M_\pi^2)
+ \left[ M_\pi^2-\omega^2 \right]
J_0(\omega) \right) \; .
\label{Sigmaab}
\end{eqnarray}
The functions ${1 \over i}\Delta $ and $J_0$ are standard one-loop integrals
given in Appendix B.
The contact contributions of Fig. \ref{self} c) have the simple form
\begin{equation}
\Sigma_{\rm contact}=-{k^2\over 2 m} +\Sigma_{\rm CT}^{(2)}+
\Sigma_{\rm CT}^{(3)}
\label{Sigmacontact}
\end{equation}
where
\begin{eqnarray}
\Sigma_{\rm CT}^{(2)}&=&-{8 \hat{m}^2 \over m}
\left(2 f_{13}+f_{14}+ f_{15}\right) \nonumber \\
\Sigma_{\rm CT}^{(3)}&=&-{16 \hat{m}^3 \over m^2}
\left( g_3+2 g_4+4 g_5+2 g_6 \right) \nonumber\\
& & -{8 \hat{m}^2 B  \over m^2}
\left(2 \tilde{c}_{13}+\tilde{c}_{14}+ \tilde{c}_{15}\right)\, .
\label{SigmaCT}
\end{eqnarray}
The loop contributions to the selfenergy contain divergences which,
for physical quantities, can always be absorbed by a appropriate
renormalization of the the counterterms $g_i$ and $\tilde{c}_i$ in
$\Sigma_{\rm CT}^{(3)}$. This is discussed in subsection 3.2 below.

The pole-part of $S_{+-}$, $S_{-+}$ and $S_{--}$ is proportional to
$S_{++}$. To the order we are working we have
\begin{eqnarray}
S_{+-}(k)&=&{1 \over 2 m} P_v^+ S_{++} \left(
1-{v\cdot k +\sigma_0 \over 2 m} + {\cal O}(p^3) \right)
\not\!{k}^\bot P_v^- \nonumber\\
S_{-+}(k)&=&{1 \over 2 m} P_v^- S_{++} \left(
1-{v\cdot k +\sigma_0 \over 2 m} + {\cal O}(p^3) \right)
\not\!{k}^\bot P_v^+ \nonumber\\
S_{--}(k)&=&{1 \over (2 m)^2} P_v^-
\not\!{k}^\bot S_{++} \not\!{k}^\bot P_v^- +{\cal O}(p^3) \; .
\label{S+-}
\end{eqnarray}
Due to the presence of the projection operators $P_v^-$ the term in
proportion to $S_{--}$ is suppressed by one additional power of $p$
and does not contribute at the order we are considering here.

Following again ref. \cite{EM97} we now define the on-shell nucleon
momentum $p_N$ as
\begin{eqnarray}
p &=&   m v+k \equiv p_N +\lambda r \nonumber\\
p_N &\equiv & m_N v+Q \; .
\label{defpN}
\end{eqnarray}
The arbitrary four-vector $r$ controls the on-shell limit
$p\rightarrow p_N$ by letting the real parameter $\lambda$ tend to
zero. We can choose $r=v$ for convenience. The nucleon mass has the
expansion
\begin{equation}
m_N=m + \sigma_0 + \delta m^{(2)} \; .
\label{mN}
\end{equation}
Here we have displayed explicitly the leading order correction
$\sigma_0$. As emphasized previously we count $\sigma_0$ as a quantity
of order $p$. The remainder, $\delta m^{(2)}$, is of ${\cal O}(p^2)$ by
definition. Then we have
\begin{equation}
p_N^2=m_N^2 \quad \Longrightarrow \quad 2 m_N v\cdot Q+Q^2=0
\end{equation}
and
\begin{equation}
v\cdot k=\sigma_0+\delta m^{(2)}-{Q^2\over 2 m} + \lambda \; .
\label{vdotk}
\end{equation}
We observe that on-shell ($\lambda\rightarrow 0$)
\begin{equation}
v\cdot k-\sigma_0 = {\cal O}(p^2) \; ,
\label{Op2}
\end{equation}
which will be crucial for the analysis to follow.
\footnote{The propagator in GHBChPT depends also only on the
combination $v\cdot k-\sigma_0$. When doing loop calculations, this
can be used to show that, to leading order, the shift $\sigma_0$ in the
propagator has no net effect. This result could be obtained
alternatively by choosing a shifted mass $m \rightarrow m+\sigma_0$ in
the exponential factor occurring in the definition of the heavy baryon
field, {\it c.f.} Eq. (\ref{vdf}). However, we prefer to keep the
mass correction $\sigma_0$ explicit, since it depends on the light
quark mass.}

\subsection{The nucleon mass to ${\boldsymbol{{\cal O}(p^3)}}$}

The pole of the nucleon propagator is entirely determined by
$S_{++}$. On-shell, and to ${\cal O}(p^3)$, we have
\begin{eqnarray}
S_{++}^{(-1)}(k) &=& v \cdot k-\sigma_0-\Sigma(k) \\
&=& {1 \over 2 m} \left( p^2-m^2-
2 m \left[\sigma_0+\Sigma_{\rm CT}^{(2)}+\Sigma_{\rm CT}^{(3)}+
\Sigma_{\rm loop}(v\cdot k-\sigma_0) \right]\right) \nonumber
\label{S++result}
\end{eqnarray}
Expanding $\Sigma_{\rm loop}$ according to
\begin{equation}
\Sigma_{\rm loop}(v\cdot k-\sigma_0)=\Sigma_{\rm loop}(0)+(v\cdot
k-\sigma_0) \Sigma_{\rm loop}'(0)+...
\end{equation}
we observe that due to (\ref{Op2}) only the leading term in this
expansion has to be kept. Thus
\begin{equation}
m_N^2=m^2+2 m \left[\sigma_0+\Sigma_{\rm CT}^{(2)}+\Sigma_{\rm CT}^{(3)}+
\Sigma_{\rm loop}(0) \right] \; .
\label{mN2result}
\end{equation}

The next step consists of removing the divergences in
$\Sigma_{\rm loop}(0)$ by renormalization of the counter term coupling
constants. Employing (\ref{Sigmaloop})
the divergent part of the one-loop selfenergy can be written as
\begin{equation}
\Sigma_{\rm loop}(0)_{|\ {\rm div}}={24 e_1 \over F^2} \left(
B \hat{m}^2+2 A \hat{m}^3 \right) \cdot \Lambda(\mu)
\end{equation}
where $\Lambda(\mu)$ contains the pole for $d\rightarrow 4$ and is
defined in Appendix B. Defining renormalized couplings
\begin{eqnarray}
g_i=g_i^r(\mu)+{m^2\over F^2} \beta_{g_i} \cdot \Lambda(\mu) \nonumber\\
\tilde{c}_i=\tilde{c}_i^r(\mu)+{m^2\over F^2} \beta_{\tilde{c}_i}
\cdot \Lambda(\mu)
\label{reng}
\end{eqnarray}
the nucleon mass in Eq. (\ref{mN2result}) is rendered finite for
\begin{eqnarray}
2 \beta_{\tilde{c}_{13}}+\beta_{\tilde{c}_{14}}
+\beta_{\tilde{c}_{15}}
-3 e_1 &=& 0
\nonumber\\
\beta_{g_3}+2 \beta_{g_4} +4 \beta_{g_5} +2 \beta_{g_6} -3 e_1 A &=& 0
\; .
\label{betag}
\end{eqnarray}
The constants $f_i$ need not to be renormalized.

Finally we solve Eq. (\ref{mN2result}) for $m_N$. Neglecting
consistently higher order terms we find
\begin{equation}
m_N=m+\Delta m^{(1)}+\Delta m^{(2)}+\Delta m^{(3)} + {\cal O}(p^4)
\label{mNresult}
\end{equation}
with
\begin{eqnarray}
\Delta m^{(1)} &=& \sigma_0 \nonumber\\
\Delta m^{(2)} &=& -{8 \hat{m}^2 \left( 2 f_{13}+f_{14}+f_{15} \right)+
{1\over 2} \sigma_0^2 \over m_N} \nonumber \\
\Delta m^{(3)} &=& -{16 \hat{m}^3 \over m_N^2} \left( g_3^r+2 g_4^r+
4 g_5^r+2 g_6^r \right) \nonumber\\
& & -{8 \hat{m}^2 B \over m_N^2} \left(
2 \tilde{c}_{13}^r+\tilde{c}_{14}^r+\tilde{c}_{15}^r \right)
\nonumber\\
& & -{3 \sigma_0 M_\pi^2 \over 32 \pi^2 F_\pi^2} \ln {M_\pi^2 \over
\mu^2}-{3 \gas \over 32 \pi F_\pi^2} M_\pi^3 \; .
\label{Deltami}
\end{eqnarray}
We have expressed everything in terms of the physical nucleon mass
$m_N$. In standard HBChPT only $\sigma_0$ and the term in proportion
to $M_\pi^3$ enter at ${\cal O}(p^3)$. The additional terms in
proportion to $f_i$, $g_i$ and $\tilde{c}_i^r$ are not known and prevent us
from giving a quantitative estimate of the nucleon mass shift.\footnote{Only
the combinations $f_i+{B\over m_N} \tilde{c}_i^r$,
$i=13,14,15$ enter
the expression and can possibly be fixed from experiment.}
However, the formula can be used in further applications of
GHBChPT and we hope that the unknown constants can be determined
in such future work.

\subsection{Wavefunction renormalization to ${\boldsymbol{{\cal O}(p^3)}}$}

The wavefunction renormalization constant $Z_N$ is defined via
\begin{eqnarray}
Z_N(Q) u(p_N) &=& \lim_{p\rightarrow p_N} S_N(p) (\not\!{p}-m_N)
u(p_N)
\nonumber\\
&=& \lim_{\lambda\rightarrow 0} S_N(p) \lambda \not\!{v} u(p_N) \; .
\end{eqnarray}
Employing (\ref{vdotk},\ref{Op2}) and keeping terms linear in $\lambda$ only
we may expand
\begin{equation}
\Sigma_{\rm loop}(v \cdot k-\sigma_0)=\Sigma_{\rm loop}(0)+
\lambda \Sigma_{\rm loop}'(0) +{\cal O}(\lambda^2,p^4) \; .
\end{equation}
Thus
\begin{equation}
Z_N(Q) u(p_N)={m \left( P_v^+ +{1\over 2 m} \not\!{k}^\bot
\left[ 1-{v\cdot k+\sigma_0 \over 2 m}\right] \right) \not\!{v} u(p_N)
\over
v\cdot p_N-m \Sigma_{\rm loop}'(0) } \; .
\end{equation}
The same steps as performed in \cite{EM97} but keeping track of the
peculiarities due to \mbox{GHBChPT} lead to the final result
\begin{equation}
Z_N(Q)=1-{1\over m_N} \left(\sigma_0+\Delta m^{(2)}\right)
+{Q^2 \over 4m_N^2}+ \Sigma_{\rm loop}'(0)
\end{equation}
where
\begin{equation}
\Sigma_{\rm loop}'(0)=-{9 \over 2} {\gas M_\pi^2 \over 16 \pi^2
F_\pi^2} \left( 16 \pi^2 \Lambda(\mu) +\ln {M_\pi \over \mu}+{1\over
3} \right) \; .
\end{equation}
This result is identical to the result obtained in standard HBChPT
except for the appearance of the term in proportion to $\Delta
m^{(2)}$. The fact that the wavefunction renormalization ``constant''
depends on the momentum $Q$ is not affected.
As before we have expressed all contributions in terms of the
physical nucleon mass.

\section{The scalar form factor of the nucleon}

As a further application of our formalism we consider the scalar
form factor of the nucleon
\begin{equation}
\bra \Psi(p^{\prime}) | \hat m\,\left(\bar{u} u+ \bar{d} d
\right)|\Psi(p)\ket = \sigma (t) \bar u(p^{\prime}) u(p) \, ,
\label{sigmadef}
\end{equation}
which is a measure of explicit chiral symmetry breaking due to up and
down quark mass.
The variable  $t=(p^{\prime}-p)^2$ denotes the square of the
four-momentum transfer.
At $t=0$ the scalar form factor yields the so called sigma term of the
nucleon, which has attracted much attention over the years.
The interest in this quantity derived partly from the discrepancy
between early determinations of $\sigma(0)$ from $\pi N$-scattering
data and naive estimates based on the analysis of the baryon mass
spectrum. Gasser, Leutwyler and Sainio have resolved the issue by a
thorough dispersive analysis of the nucleon sigma term. \cite{GLS91a}
The method relies on three steps: i) a low energy
theorem due to Brown, Pardee and Peccei \cite{BPP71} relates the
isospin even $\pi N$-scattering amplitude (with Born term removed) at
the Cheng-Dashen point, $\bar{D}^+(2 M_\pi^2)$, to the scalar form
factor of the nucleon, {\it i.e.}
\begin{equation}
\Sigma\equiv F_\pi^2 \bar{D}^+(2 M_\pi^2)=\sigma(2 M_\pi^2) +
\Delta_{\rm R} \; .
\label{LET}
\end{equation}
The remainder $\Delta_{\rm R}$ is of ${\cal O}(\hat{m}^2)$ and was
recently calculated to order $p^4$ in standard HBChPT. \cite{BKM96}
It was shown that potentially large contributions
due to chiral logarithms of the form $M_\pi^4 \ln M_\pi$ cancel
exactly; $\Delta_{\rm R}$ is indeed small and was estimated to be
bounded by 2 MeV. ii) $\Sigma$ is determined by the extrapolation of the
$\pi N$ scattering amplitude from the physical region $t\leq 0$ to the
Cheng-Dashen point. It is useful to decompose
\begin{equation}
\Sigma=\Sigma_{\rm d}+\Delta_{\rm D}\, ,
\end{equation}
where $\Sigma_{\rm d}$ denotes the first two terms in a Taylor series
expansion. The point is that $\Sigma_{\rm d}$ is fixed in terms of the
$\pi N$ scattering amplitude in the physical region. The remainder,
$\Delta_{\rm D}$, is also accessible through a dispersive analysis but
depends on the $\pi\pi$ phase shift. It is particularly sensitive to
the region just above the two pion threshold. Numerically, the
analysis in \cite{GLS91b} yielded $\Sigma_{\rm d}=48$ MeV, with an
error bar of about 8
MeV, and $\Delta_{\rm D}=11.9\pm 0.6$ MeV.
We refer to \cite{GLS91a,GLS91b,Sai94} for a detailed account
on this part. iii) In order to determine the nucleon sigma term from
(\ref{LET}), it remains to calculate the shift of the scalar form
factor from the Cheng-Dashen point to zero momentum transfer,
\begin{equation}
\Delta_{\sigma} = \sigma (2\mpi) - \sigma (0) \, .
\end{equation}
In \cite{GLS91b} $\Delta_{\sigma}$ was calculated by means of a once
subtracted dispersion relation for $\sigma(t)$, yielding
\begin{equation}
\Delta_{\sigma} = 15.2 \pm 0.4 \; \text{MeV} \, .
\end{equation}
It was noted that $\Delta_\sigma$ cannot be reliably calculated in
chiral perturbation theory, say to one-loop. The reason is that the
imaginary part entering the dispersion relation is in proportion to
$\Gamma_\pi(t)=\bra \pi^0(p')|\hat{m}(\bar{u} u+\bar{d}
d|\pi^0(p)\ket$, the scalar form factor of the pion, and to
$f_+^0$, the I=J=0 $\pi N$ partial wave in the t-channel. However, both
are grossly underestimated in the two-pion threshold region if
leading order ChPT calculations are employed.

If the quark condensate is substantially smaller than assumed in
standard ChPT, all of the three steps mentioned above are subject to
modifications and must be reanalyzed. In step ii) and iii), the
$\pi\pi$ phase shift close to threshold plays an important role.
Moreover, the normalization of the pion scalar
form factor, which enters the determination of $\Delta_\sigma$,
is also sensitive to the light quark condensate.
\footnote{We are grateful to Jan Stern for pointing this out to us.}
This can be seen best by using the
Feynman-Hellman theorem for $\Gamma_{\pi}(0)$, {\it i.e.}
\begin{equation}
\Gamma_{\pi}(0)  =\, \hat m \frac{\partial \mpi}{\partial \hat m}
             =\, 2B\mh +8A\mh^2\equiv \mpi (2-x)\; ,
\label{normalization}
\end{equation}
with
\begin{equation}
x\equiv x_{\text{GOR}} =\, \frac{2B\mh}{\mpi}\, ,\qquad  0\leq x \leq\, 1 \,
.
\end{equation}
The parameter $x$ interpolates between the extreme generalized limit, $x=0$,
and the standard case with $x=1$. Note that $x$ is given in terms of $B$, a
quantity of chiral $SU(2)$. The relation between $x$ and the more familiar
ratio $r=m_s/\hat{m}$ is given in \cite{KMSF96}. For $r\lesssim 12$ the
normalization of the pion scalar form factor starts to deviate
strongly from the standard case. We will come back to a dispersive
treatment of $\sigma(t)$ in subsection 4.2 but now turn to the
GHBChPT calculation.

\subsection{Scalar form factor of the nucleon to ${\cal O}(p^3)$ in GHBChPT}

\begin{figure}[p]
\begin{center}
\leavevmode
\epsfig{file=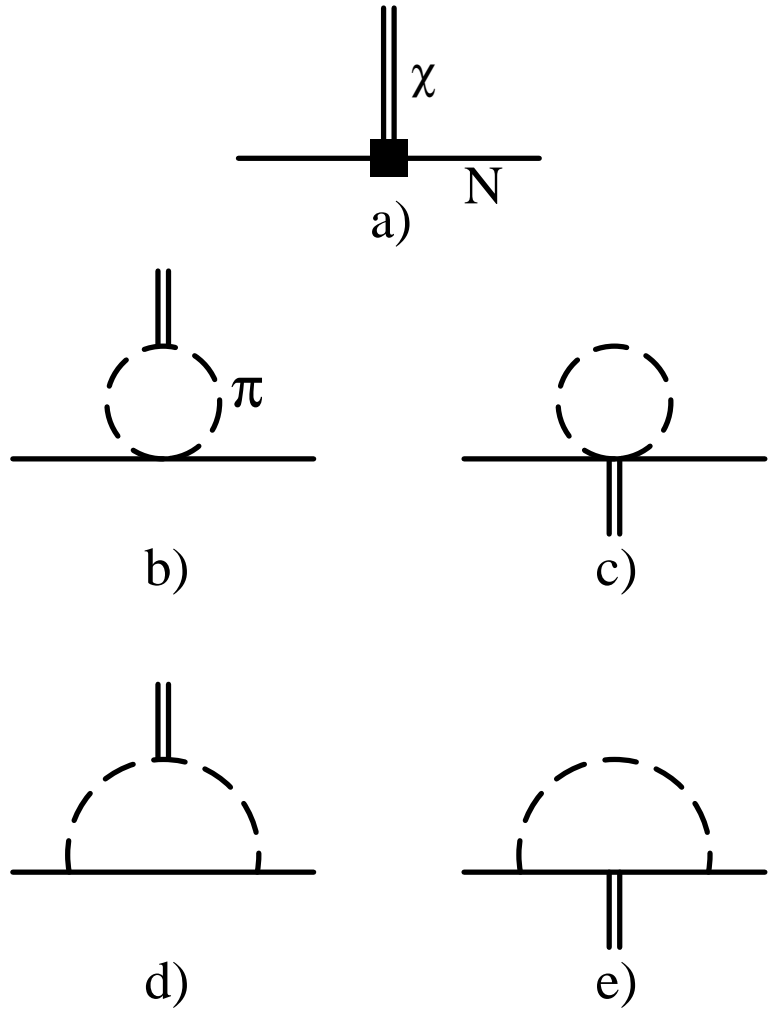,width=10cm}
\caption[]{Feynman diagrams contributing to the scalar form factor of the
nucleon to ${\cal O}(p^3)$.  Plain,  dashed, and double lines line
denote the nucleon, the  pion, and the scalar source, respectively.
The  shaded box  denotes tree contributions of order $p$, $p^2$, and
$p^3$.}
\label{graphs}
\end{center}
\end{figure}

Although it is clear from the above that ChPT itself cannot provide a
full understanding of the nucleon sigma term, ChPT is nevertheless needed
in order to provide important constraints on the dispersive analysis.
The method of calculating fully relativistic quantities like the
scalar form factor in Eq. (\ref{sigmadef}) by using HBChPT has been
spelled out in \cite{EM97}. We follow this method and also employ the
initial nucleon rest frame (INRF) when applying wave function
renormalization. The fully renormalized scalar
form factor thus calculated in HBChPT coincides with the relativistic form
factor we are seeking. The Feynman diagrams contributing to order
$p^3$ GHBChPT are shown in Fig. \ref{graphs}.  Compared to the standard case
there are three additional loop diagrams, {\it i.e.} graphs b), c) and e).
In dimensional regularization the sum of all loop graphs is found to be
\begin{equation}\label{loop}
\begin{split}
\sigma(t)_{\text{loop}} =&\, -\frac{3}{2}\sigma_0
                     \frac{\Delta (M^{2}_{\pi})}{F^2}
+ \frac{3\gas\sigma_0}{4F^2} \bigl[
                   \Delta (M^{2}_{\pi}) -M^{2}_{\pi}J^{'}_{0}(0)\bigr] \\
&\,+\frac{3}{2}\left(2-x\right)\sigma_0\frac{\mpi}{F^2}
                J_{\pi\pi}(t)
-\frac{3}{8}\left(2-x\right) \frac{ \gas  \mpi }{F^2} \left[ \bigl(
t-2M^{2}_{\pi}) K_0 (0,t) -2 J_0(0) \right]
\, ,\\
\end{split}
\end{equation}
  where  $\sigma_0$ is the lowest order contribution to $\sigma(t)$
\begin{equation}
\sigma_0=  -4\hat m e_1 \, .
\end{equation}
The loop functions $\Delta$, $J^{'}_{0}$, $J_{\pi\pi}$, $K_0$, and
$J_0$ are given in Appendix \ref{appendixc}.

The tree-contributions in Fig. \ref{graphs} a) are modified as well. Besides
the ${\cal
O}(p)$ contribution, which in the standard case was counted as ${\cal
O}(p^2)$, there are genuine new vertices of the order $p^2$ and
$p^3$. These can be read off easily from the corresponding effective
lagrangians given in section 2. We find
\begin{equation}
\begin{split}
\sigma_{\text{tree}}(t) = &\, \sigma_0
             -\frac{16{\mh}^2}{m} \left[ 2f_{13}+f_{14}+ f_{15}
           +\frac{B}{m}\left( 2 \tilde{c}_{13}+\tilde{c}_{14}+
\tilde{c}_{15}\right)
                \right]              \\
&\,-\frac{48{\hat m}^3}{m^2} \left( g_3 +2g_4+4g_5+2g_6 \right)
+\frac{\hat m t}{m^2} \left( g_1 +2g_2 \right)\, .\\
\end{split}
\end{equation}

Finally we must take into account wave-function renormalization of the
in- and outgoing fields $\nv$ and $\nvb,$ respectively,
\begin{equation}
\begin{split}
\sigma (t) &\,=(\sigma_{\rm tree}(t)+\sigma_{\rm loop}(t) )\sqrt{Z_{N}(0)
Z_{N}(Q)}\; .\\
\end{split}
\end{equation}
As mentioned above, we work in the INRF where incoming and outgoing
nucleon four momentum are given as
\begin{equation}
p_{\rm in}=m_N v, \qquad p_{\rm out}=m_N v+Q  \, , \quad Q=p-p^{\prime} \; .
\end{equation}

For the following it is convenient to decompose the scalar form factor
according to
\begin{equation}
\sigma(t)\equiv \sigma(0)+\bar{\sigma}(t)
\end{equation}
and to discuss the two pieces separately. The contributions to
$\sigma(0)$ of the
loop-graphs as well as those from wavefunction renormalization
 contain divergences for $d\rightarrow 4$. These are
removed by introducing renormalized coupling constants $g_i^r$ and
$\tilde{c}_i^R$ as given in Eqs. (\ref{reng},\ref{betag}). Using the
explicit form of the loop functions at zero momentum transfer and
expanding consistently up to ${\cal O}(p^3)$, we
arrive at the final result for the nucleon sigma term
\begin{eqnarray}
\sigma(0)&=& \sigma_0 +\left( 2-{\sigma_0 \over m_N} \right) \Delta
m^{(2)} \nonumber\\
& & -{16 \hat{m}^2 B \over m_N^2} \left(
2 \tilde{c}_{13}^r+\tilde{c}_{14}^r+\tilde{c}_{15}^r \right)
\nonumber\\
& & -{48 \hat{m}^3 \over m_N^2} \left( g_3^r+2 g_4^r+
4 g_5^r+2 g_6^r \right) \nonumber\\
& & -{3 \sigma_0 M_\pi^2 \over 32 \pi^2 F_\pi^2} \left[
(3-x) \ln {M_\pi^2 \over \mu^2}+(2-x)\right]-
{9 \gas \over 64 \pi F_\pi^2} (2-x) M_\pi^3 \; ,
\label{sigma0final}
\end{eqnarray}
where $\Delta m^{(2)}$ was given in (\ref{Deltami}).
This result agrees with the Feynman-Hellman theorem
\begin{equation}
\sigma(0)=\hat{m} {\partial \over \partial \hat{m}} m_N
\end{equation}
and therefore provides a nice check on our calculation.

The $t$--dependent part of the scalar form factor involves only
finite loop functions and needs no infinite renormalization of
counterterms. We thus obtain the scale independent result
\begin{eqnarray}
\bar \sigma(t) &=& {\sigma_0+ 8 \hat{m} (g_1+2 g_2) \over 8 m_N^2}
\cdot t +{3 \sigma_0 \mpi \over 2 F_\pi^2} (2-x) \bar J_{\pi\pi}(t)
\nonumber \\
& & -{3 \gas \mpi \over 8 F_\pi^2} (2-x) \left[ (t-2 \mpi) K_0(0,t)-
{M_\pi \over 8 \pi} \right] \; .
\label{sigmatfinal}
\end{eqnarray}
The loop contributions scale with $(2-x)$, {\it i.e.} there
is a factor of two difference between the extreme generalized $(x=0)$ and
standard case $(x=1)$. The polynomial part linear in $t$ on the other
hand does not exhibit this scale factor. The combination of coupling
constants $g_1+2 g_2$ is unknown; in standard ChPT these terms would
occur at order $p^4$.

The phenomenological implications of these results can be assessed only
in comparison with the dispersive analysis as described in
\cite{GLS91a,GLS91b} for the standard case and outlined at the
beginning of this section. As to $\sigma(0)$, we do not know the
coupling constants $e_1$, $\tilde{c}^r_i$, and $g_i^r$. On dimensional
grounds, we expect these constants to be of order unity. The leading
order term, $\sigma_0$, can be estimated as follows. The counter term 
contributions of ${\cal O}(p^2)$ are suppressed by additional factors 
$\hat{m}/m_N$ and therefore must be small. The loop contributions, 
{\it i.e.} the last line in
Eq. (\ref{sigma0final}), depend explicitly on the ratio $x$. The term in
proportion to $\gas$ is dominating and numerically yields $-(2-x) 22.5$ MeV.
Once the sigma term is determined from a dispersive analysis, we then have
\begin{equation}
\sigma_0 \approx \sigma(0)^{\rm dispersive}+ (2-x)\,  22.5\,  {\rm MeV}.
\label{sigma0fix}
\end{equation}
Note that the result of the dispersive analysis will also depend on $x$.
Incidentally, this variation with $x$ partially cancels the $x$-dependence
of the second term in Eq. (\ref{sigma0fix}), leading to 
$\sigma_0 \approx 67.5 \ldots 80$ MeV, where the lower and upper bound 
correspond to $x=1$ and $x=0$, respectively. \cite{KB98} 

The phenomenological analysis of the $t$-dependent part is also instructive.
Consider the shift between Cheng-Dashen point and zero momentum
transfer to ${\cal O}(p^3)$
\begin{equation}
\begin{split}
\Delta_{\sigma }  &\, =  \frac{2 M^3_{\pi} }{m^2_N} \left[
\frac{\mh}{M_{\pi}}\left(g^r_1 +2g^r_2\right)
+\frac{\sigma_0}{8M_{\pi}}
\right]
+\left(2-x\right)\frac{2M^3_{\pi}}{(4\pi F) ^2} \frac{3\pi}{8} \left[
\gas +\frac{\sigma_0}{M_{\pi}} \left( \frac{4}{\pi}-1 \right) \right]
 \, .
\end{split}
\end{equation}
The second term in the first square bracket is due to 
wave-function renormalization. The standard result to order
${\cal O}(p^3)$ is obtained by
setting $x=1$ and by the observation that in this case all terms
in proportion to $\sigma_0$ and the couplings $g^r_i$ are at least of
${\cal O}(p^4)$.
 Setting $\ga=1.26$, $F_{\pi}=92.4$ MeV, $m_N=939$ MeV we obtain
\begin{equation}
\Delta_{\sigma }=\left\{ 6\left[\frac{\mh}{M_{\pi}}\left(g^r_1 +2g^r_2\right)
+\frac{\sigma_0}{8M_{\pi}}
\right]+(2-x) \left[ 7.5 +1.6\frac{\sigma_0}{M_{\pi}}
\right] \right\}
\text{MeV} \, .
\label{delsigmaestimate}
\end{equation}

In order to get a rough estimate of the size of the contribution not in
proportion to $(2-x)$
we set $\sigma_0=70$ MeV and employ a typical value for the light
quark mass in generalized ChPT, $\hat{m}=20$ MeV. On dimensional analysis
grounds one expects the coupling constants $g^r_1$ and $g^r_2$ to be of
order unity. Varying the sum $g^r_1 +2g^r_2$ between the bounds
$\pm 3$   yields a net contribution of about $\pm 3$ MeV to
$\Delta_{\sigma }$. The second term in Eq. (\ref{delsigmaestimate}) is due
to finite loop graphs. Numerically, the term in proportion to $\gas$ 
is the dominant contribution
for reasonable values of $\sigma_0$, i.e. $\sigma_0=60 \dots 80$ MeV.
Due to the scale factor $(2-x)$ the shift $\Delta_\sigma$ 
yields contributions which 
can be larger than the standard ChPT result to ${\cal O}(p^3)$ by more 
than a factor of 2.

How does this result compare to a dispersive analysis
adopted to the case of a small quark condensate? While a full
treatment is outside the scope of this article, we sketch in
subsection 4.2 below a dispersive treatment of $\bar{\sigma}(t)$.
Details are deferred to a forthcoming publication. \cite{KB98}
In Fig. \ref{imaginary} we compare the imaginary part
of $\sigma(t)$ of the dispersive analysis with the result of the
${\cal O}(p^3)$ GHBChPT calculation. The dispersive result is seen
to yield a strong enhancement over the ChPT calculation, as pointed
out in \cite{GLS91b}. The figure clearly shows that a leading order
ChPT calculation is not appropriate to calculate the
imaginary part reliably, both in the standard as well as in the
generalized case. Note that the universal factor $(2-x)$ has been
divided out. The comparison also shows that the failure of the GHBChPT
calculation for $\Delta_\sigma$ (compare Table 1) cannot be blamed on
the dimensional
estimate for the coupling constant $g_1^r+2 g_2^r$. We know that
higher order corrections in the chiral expansion will modify the
imaginary part substantially. Neglecting these and adjusting
$g_1^r+2 g_2^r$ such that the dispersive value for $\Delta_\sigma$ is
reproduced is therefore without justification.

\subsection{Dispersive analysis of $\bar{\sigma(t)}$}

We present a dispersive analysis of $\bar{\sigma}(t)$
adopted to the case of a small quark condensate. The model serves as
a representation to which the ChPT calculation can be compared, but also
provides a first step towards the calculation of the nucleon sigma
term. We employ the once subtracted dispersion relation
\begin{equation}
\bar{\sigma}(t)={t\over \pi} \int dt' {{\rm Im}\, \sigma(t') \over t'
(t'-t-i
\epsilon) }
\label{disprelation}
\end{equation}
and use the imaginary part as given in the elastic region $4 M_\pi^2 <
t <16 M_\pi^2$ via
\begin{equation}
{\rm Im}\, \sigma(t)={3\over2} {\Gamma_\pi^*(t) f_+^0(t)\over 4 m_N^2-t}
\left( 1-{4 M_\pi^2 \over t}\right) ^{1\over 2}\; .
\label{Imsigma}
\end{equation}
Here, $\Gamma_\pi(t)$ is the scalar form factor of the pion and $f_+^0(t)$
is the I=J=0 $\pi N$ partial wave in the t-channel. Both of these
amplitudes are subject to strong final state interactions of the two
pions. In order to account for these effects, we model the imaginary
part as follows. For the pion form factor, we use the form
\begin{equation}
\Gamma_\pi(t)=\Gamma_\pi(0) (1+b\cdot t) \exp\{ \Delta_0(t)\} \; ,
\label{Gammadisp}
\end{equation}
where
\begin{equation}
\Delta_0(t)={t\over \pi} \int_{4 M_\pi^2}^{t_1} dt' {\delta_0^0(t')
\over t' (t'-t-i \epsilon)}
\label{Omnes}
\end{equation}
is referred to as the Omn\`es function and $\delta_0^0$ denotes the
I=J=0 $\pi\pi$ phase shift. For definiteness, the cutoff in
(\ref{Omnes}) is taken as $t_1=(0.9\, {\rm GeV})^2$.

\begin{table}
\begin{center}
\setlength{\extrarowheight}{10pt}
\begin{tabular}{|l|cccccc|}
\hline
$\alpha $    & 1 & 2 & 2.5 & 3 & 3.5 & 4  \\
\hline
$r={m_s\over \hat{m}}$    & $\approx 25$ & 12 & 10.25 & 9.2 & 8.5 &
$\approx 8$ \\
$2-x$                    & 1    & 1.40 & 1.58 & 1.75 & 1.93 & 2.00 \\
$\Delta_\sigma^{\rm gen}/\Delta_\sigma^{\rm stan}$
                         & 1    & 1.38 & 1.55 & 1.71 & 1.87 & 1.93 \\
${d\sigma^{\rm gen}(0)\over dt} /
 {d\sigma^{\rm stan}(0)\over dt }$
                         & 1    & 1.37 & 1.53 & 1.68 & 1.84 & 1.89 \\
[0.15cm]
\hline
\end{tabular}
\caption{$r$-dependence of $\Delta_\sigma$ and ${d\sigma(0) \over dt}$
normalized to the standard case}
\end{center}
\end{table}

The generalized
scenario of SB$\chi$S enters Eq. (\ref{Gammadisp}) in two ways. First,
the normalization $\Gamma_\pi(0)$ is sensitive to the quark
condensate, and to leading order it is given by Eq. (\ref{normalization}).
Higher order corrections are expected to be small and we employ the
simple form (\ref{normalization}) in our dispersive analysis. Second,
the I=J=0 $\pi\pi$ phase shift differs considerably from the standard
case, in the threshold region. Consequently, the Omn\`es function is
modified, leading to a further enhancement of the pion form factor
close to threshold. The polynomial $(1+b\cdot t)$ in (\ref{Gammadisp}) is
used to mimic the effect of a two-channel analysis which, in the
standard case, was seen to
modify the pion form factor substantially above $t=(0.45\, {\rm GeV})^2$.
\cite{DGL90}. We expect these effects to a large extent to be 
independent of the $\pi\pi$ phase shift at threshold and fix the
parameter universally at $b=0.038\, {\rm fm}^2$.

As to $f_+^0$ we employ the strategy developed in \cite{GLS91b,FF60}
in order to extrapolate to the unphysical region. For $t\leq 0$ we use
the data tabulated in \cite{Hoe83}. The continuation depends on the
$\pi\pi$ phase shift employed -- here enters the assumption made about
the quark condensate. We use a parameterization of $\delta_0^0$ given
by Schenk \cite{Sch91} with threshold parameters $a_0^0$, $b_0^0$
taken from \cite{KMSF96}. Details of this analysis will be presented
\mbox{elsewhere \cite{KB98}.}

\begin{figure}
\begin{center}
\leavevmode\epsfig{file=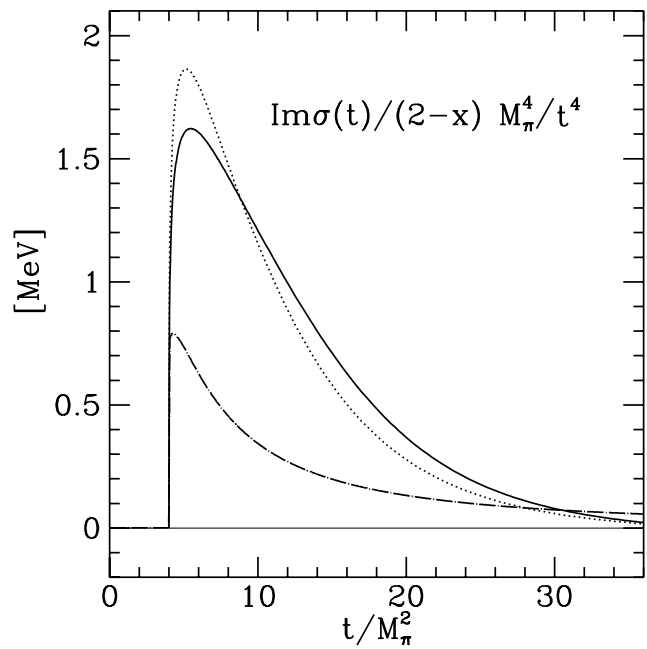,width=9cm}
\caption[]{The imaginary part of the scalar form factor. Full, dotted
and dashed-dotted lines denote the dispersive analysis
in the standard case, the dispersive analysis in the extreme
generalized case and the ${\cal O}(p^3)$ GHBChPT calculation, respectively.}
\label{imaginary}
\end{center}
\end{figure}

The model specified by Eqs. (\ref{disprelation}-\ref{Omnes}) and the
continuation of $f_+^0$ just described is
then used to calculate $\bar{\sigma}(t)$ for $0<x<1$.
In Fig. \ref{imaginary} the imaginary parts of $\sigma(t)$ for the two
limiting
cases are compared.
The threshold enhancement of the curve
corresponding to the dispersive treatment of the extreme generalized
case ($x=0$) over that of the standard case ($x=1$) is due to the
larger scattering length of the former. The area
under the two dispersive curves in Fig. \ref{imaginary} is roughly equal,
however.
The ratio $\Delta_\sigma^{\rm generalized}(x) / \Delta_\sigma^{\rm
standard}$ is thus close to $(2-x)$. In Table 1 we
give this ratio as well $d \sigma(0)/dt$ normalized to the standard
case for
various values of $x$. $\alpha$ is related to the $\pi\pi$ scattering
amplitude at the symmetric point $s=t=u$. \cite{stern91} For the sake of
comparison we also give the corresponding values of $r$, taken from
\cite{KMSF96}, and $2-x$.
The shift of scalar form factor of the nucleon deviates from the
standard case substantially for $r\lesssim 12$.

We close the discussion with
a remark concerning the dispersive analysis of the nucleon sigma
term itself. As mentioned at the beginning of this
section, $\sigma(0)$ can be obtained by updating the estimates for
$\Delta_{\rm R}$ and $\Delta_{\rm D}$, adopted to the case of a small
quark condensate. $\Delta_{\rm D}$ is particularly sensitive to the
threshold
behaviour of the $\pi\pi$ phase shift. The
crucial question here is whether the almost perfect cancellation
between the remainders $\Delta_{\rm D}$ and $\Delta_\sigma$ observed in
\cite{GLS91a} persist in the case of a small quark condensate.
Work in this direction is under way and results will be reported in
\cite{KB98}.

\section{Conclusions}

The consequences of a small quark condensate are studied for the
baryonic sector of ChPT. To this end, we have constructed an effective
theory of the $\pi N$--system respecting chiral symmetry and
admitting a systematic expansion in small momenta, the light quark
masses, and the dimensionful parameter $B=-<\bar q q>/F^2$,
collectively denoted as $p$ (GHBChPT). The light quark masses are counted as
order $p$, in contrast to the standard counting rule $m_q \sim p^2$.
Moreover, we assume that in the chiral limit the theory contains no
other small scales than $B$. The effective lagrangian is given in
it's most general form to ${\cal O}(p^2)$ and to ${\cal O}(p^3)$
in the scalar sector. A method to efficiently construct the relativistic
baryonic chiral lagrangians for chiral SU(2) to all orders is given in
the Appendix.

Mass- and  wavefunction renormalization have been calculated to ${\cal
O}(p^3)$. These results will be useful for future applications of the
formalism laid out in this article. We have, then, considered the
scalar form factor of the nucleon to order $p^3$. The result depends
on additional low energy coupling constants not present in the
standard case at this order. By comparison to a dispersive treatment
of the subtracted nucleon scalar form factor adopted to the generalized
scenario of SB$\chi$S, it is shown that the chiral prediction for the
shift $\Delta_\sigma=\sigma(2 M_\pi^2)-\sigma(0)$ is unreliable also
in generalized ChPT. Moreover, the dispersive analysis yields a strong
deviation of $\Delta_\sigma$ from the standard result provided
$r=m_s/\hat{m}\lesssim 12$, which can reach up to a factor of two for the
limiting case of a vanishing quark condensate. In order to determine
the nucleon sigma term, both the remainder at the Cheng-Dashen point,
$\Delta_{\rm R}$, as well as the remainder in the extrapolation of the
$\pi N$ scattering amplitude from the physical region to the
Cheng-Dashen point, $\Delta_{\rm D}$, have to be reanalyzed without
the assumption of a large quark condensate.

Other processes like $\pi N$-scattering or $\pi N\rightarrow N \pi\pi$
are expected to be sensitive to the value of the light quark
condensate too. We hope that future studies in the framework presented
here will lead to a determination of many of the unknown coupling
constants. This, together with dispersive theoretic methods, should
ultimately make it possible to test the standard scenario
of spontaneous breakdown of chiral symmetry   in the baryonic
sector as well.

\begin{center}
{\bf Acknowledgments}
\end{center}
\noindent
We would like to thank M.~Knecht for correspondence and J.~Stern for
discussion initiating our work on the dispersive treatment of the
nucleon scalar form factor. This work was supported in part by
Schweizerischer Nationalfonds.

\newpage

\newcommand{\PL}[3]{{Phys. Lett.}       {\bf{#1}} {(19#2)} {#3}}
\newcommand{\PRL}[3]{{Phys. Rev. Lett.} {\bf{#1}} {(19#2)} {#3}}
\newcommand{\PR}[3]{{Phys. Rev.}        {\bf{#1}} {(19#2)} {#3}}
\newcommand{\NP}[3]{{Nucl. Phys.}       {\bf B{#1}} {(19#2)} {#3}}


\newpage
\appendix
\section{Construction of a ${\boldsymbol{SU(2)}}$ invariant Lagrangian}
\label{appendixa}
In the following we will give a recipe of how to efficiently construct the
relativistic baryon-meson Lagrangian ${\cal L}_{\pi N}$. We demand that
${\cal L}_{\pi N}$ be hermitian, flavor neutral, invariant under $SU(2)$
chiral transformations, proper Lorentz transformations, and the discrete
symmetries\footnote{ The transformation properties of the chiral fields and
of the elements of the Clifford Algebra are listed in table \ref{chiralc}
and table  \ref{cliffc} respectively.} C,P, and T. The method follows
closely that used by Krause for the case of chiral $SU(3)$ \cite{Kr90}.
There are some differences, however, which are due to the fact that in
chiral $SU(2)$ the Baryons belong to the fundamental representation. Also,
we feel an explicit exposition of the rules employed will be helpful for
future work in HBCHPT, both in the standard as well as in the 
generalized version. 
An alternative derivation has appeared recently in ref. \cite{FMS98}.

The construction of the effective Lagrangian is built upon the traceless 
chiral fields
\begin{equation}
\label{traceless}
u_{\mu}, \, f_\pm^{\mu\nu}, \, \chi_\pm -\frac{1}{2}\bra \chi_\pm\ket \, ,
\end{equation}
and the singlet chiral fields
\begin{equation}\label{singlet}
\bra \chi_\pm\ket \, , v_{\mu\nu}^{(s)} \, .
\end{equation}
All of these are $2\times 2$ matrices in flavor space.  In order to
keep the following exposition and expressions as simple as possible,
we slightly change our notation of the covariant derivative. In the
previous sections a covariant derivative   acted on all fields to the
right of it. Now it is understood that a derivate acts only on the
field right next to it, {\it i.e.}
 \begin{equation}
\nabla_{\! \mu}X=\partial_{\mu}X +\left[\Gamma_{\mu},X\right] \, .
\end{equation}
The covariant derivative on the nucleon field $\Psi$ is given by
\begin{equation}
\nabla_{\! \mu} \Psi= \partial_\mu \Psi  + \left(\Gamma_\mu - i
v_\mu^{(s)}\right)\Psi  \, .
\end{equation}
The chiral order of the  fields are:
\begin{equation}
\chi_\pm,\,\chi_\pm -\frac{1}{2}\bra \chi_\pm\ket ,\,u_{\mu}\sim {\cal O}(p)
\, , \quad
f_\pm^{\mu\nu}, \, v_{\mu\nu}^{(s)} \sim {\cal O}(p^2) \, .
\end{equation}
A covariant derivative acting on these fields increases the chiral order by
one.
The covariant derivative on the nucleon field $\Psi$ counts as
\begin{equation}
\nabla_{\! \mu} \Psi\sim {\cal O}(1) \, .
\end{equation}

In what follows we will construct the most general operator which can occur
in
${\cal L}_{\pi N}$.
In a first step we restrict ourselves  to terms in ${\cal L}_{\pi N}$
without derivatives acting on the nucleon fields $\Psi$ or $\bar \Psi$. In
this case such a term is generically of  the form
\begin{equation}
\bar \Psi \Gamma A \Psi \, ,\quad \quad\Gamma \in \text{Clifford Algebra}\,
.
\end{equation}
The term $A$ is a polynomial in the chiral fields of \eqref{traceless},
\eqref{singlet},   and
covariant derivatives thereof.
To ensure Lorentz invariance all Lorentz indices must be contracted with a
suitable combination of the metric $g^{\mu \nu}$ or the completely
antisymmetric tensor $\epsilon^{\mu \nu \rho \sigma}$.
Instead of writing the polynomial $A$ as a simple product of   chiral fields
 it is more suitable to consider the general polynomial $A$ in the somewhat
more complex form
\begin{equation}
A=\left( A_1,\left( A_2, \dotsm A_n \right)\dots\right)
\end{equation}
where $\left( A_1,A_2\right) $ denotes either the commutator
$\left[A_1,A_2\right]$ or the anticommutator $\left\{A_1,A_2\right\}$
of the chiral fields $A_1$ and $A_2$.
The polynomial A can then be simplified with the observation that
\begin{equation}
\left[ D,\left[ C, B\right] \right]  = \left\{ B,\left\{ C,D\right\}
\right\} -\left\{ C,\left\{ B, D \right\} \right\} \, ,
\end{equation}
and the fact that the anticommutator of two traceless fields $B,C \in SU(2)$
can be written as a trace in flavor space
\begin{equation}
\left\{ B,C\right\} = \bra \!  B C\ket \, .
\end{equation}
Starting with the innermost (anti-) commutator $(A_{n-1},A_n)$ it is easy to
see that the general polynomial A with $n$ non singlet chiral fields
decomposes to a linear combination of polynomials $\bar A$ of the form
\begin{equation}\label{operator}
\bar A=
\begin{cases}
\phantom{\left[ A_1, \,\, \right.}{\boldsymbol{1_d}}\,{\cal O}_n &
\\
 \phantom{\left[ A_{i_1}, \,\, \right.} A_1 \,{\cal O}_{n-1} &
\\
\left[A_{i_1},A_{i_2}\right] \,{\cal O}_{n-2} &
\end{cases}
\end{equation}
where ${\cal O}_n$ stands for a generic product of three types of flavor
traces  \begin{equation}
\begin{split}
{\cal O}_n&\,=
\bra\! A_{1} \left[ A_{2},A_{3}\right]\ket \ldots
\bra\! A_{j-2} \left[ A_{j-1},A_{j} \right]\ket
\\
&\,\times\bra \!A_{j+1}A_{j+2}\ket \ldots \bra \!A_{n-1}A_{n }\ket
\\
 &\,\times S_1 \, S_2 \ldots S_{n'}
\end{split}
\end{equation}
The operator  $A_i$ $(S_i)$  stands for a non-singlet (singlet) chiral field
and covariant derivatives thereof. For their definitions see equations
\eqref{traceless} and \eqref{singlet}. \\
Under charge conjugation such a polynomial transforms in a definite way
\begin{equation}
{\bar A}^c=(-1)^{c_{\bar A}} {\bar A}^T \, ,
\end{equation}
 where
\begin{equation}
\begin{split}
c_{\bar A} =&\,c_1+\dots +c_n+n_v +n_{[\,\,]}\, ,
 \\
n_{[\,\,]}= &\,\mbox{number of commutators in $\bar A$} \, ,
\\
n= &\,\mbox{number of non-singlet fields  in $\bar A$} \, ,
\\
n_v=&\,\mbox{number of $ v_{\mu\nu}^{(s)}$ in $\bar A$} \, ,
\end{split}
\end{equation}
and $c_k$ is the $c$-parity of the k'th chiral field $A_k$
\begin{equation}
A^{c}_{k}=(-1)^{c_k} A^T_k \, ,\quad  \text{(see table \ref{chiralc})}\, .
\end{equation}
This relation is valid because under charge conjugation the
(anti-)commutator
$(A_1,A_2)$ transforms as
\begin{equation}
(A_1,A_2)_{\mp} = \mp (-1)^{c_1+c_2} (A_1,A_2)_{\mp} \, .
\end{equation}
In a similar fashion the hermiticity property of $A$ can be analyzed
\begin{align}
{\bar A}^{\dagger}=&\,(-1)^{h_{\bar A}} \bar A\, ,\nonumber\\
h_{\bar A}  =&\, h_1+\dots +h_n +n_{\chim}+n_{[\, \,]}\, , \\
n_{\chim}=&\,\mbox{number of $ \bra\chim \ket$ in $\bar A$}  \nonumber \, ,
\end{align}
where $h_k$ is given by
\begin{equation}
A^{\dagger}_{k}=(-1)^{h_k} A_k \, ,\quad  \text{(see table \ref{chiralc})}\,
.
\end{equation}
The parity of $\bar A$ is
\begin{equation}
\begin{split}
{\bar A}^P &\,= (-1)^{p_{\bar A}} \bar A\, \\
p_{\bar A} &\, = p_1+\ldots + p_n +n_{\chim} + n_{\epsilon} \, ,\\
n_{\epsilon} &\, = \text{number of $\epsilon^{\mu \nu \sigma \rho}$ in $\bar
A$}
\, . \\
\end{split}
\end{equation}
Here, $p_k$ is the parity of the k'th chiral field $A_k$
\begin{equation}
 A^{P}_{k} = (-1)^{p_k} \bar A \, \quad \text{(see table \ref{chiralc})} \,.
\end{equation}
It is understood that under parity lower Lorentz indices in $A_k$ are
contracted
with the metric $g^{\mu \nu}$ and vice versa.

The most general term that can occur in ${\cal L}_{\pi N}$ without
 derivatives on the nucleon fields is then given by
\begin{equation}
\psib i^{\epsilon}\bar A \Psi \, , \quad \epsilon =0,1 \, .
\end{equation}
The additional factor $i $ is needed in the case when $\bar A$ is
anti-hermitian.
 We are now in the position
to derive the useful relations:
\begin{align}
\left(\bar \Psi \Gamma i^{\epsilon}\bar A \Psi\right)^c=
&    \,(-1)^{c_{\bar A}+c_{\Gamma}} \,\bar \Psi \Gamma i^{\epsilon}\bar A
\Psi
\\
\left(\bar \Psi \Gamma i^{\epsilon}\bar A \Psi\right)^{\dagger}=
& \,(-1)^{h_{\bar A} +h_{\Gamma}+\epsilon}\,\bar \Psi \Gamma
i^{\epsilon}\bar A \Psi \, ,
\\
\left(\bar \Psi \Gamma i^{\epsilon}\bar A \Psi\right)^P=
&    \,(-1)^{p_{\bar A}+p_{\Gamma}} \,\bar \Psi \Gamma i^{\epsilon}\bar A
\Psi
\end{align}
with (see table \ref{cliffc})
\begin{equation}
\left(\psib \Gamma \Psi\right)^P
     = (-1)^{p_{\Gamma}}   \psib \Gamma \Psi   \, , \quad
C^{-1} \Gamma C = (-1)^{c_{\Gamma}}\Gamma^T \, ,
\quad \gamma_0 \Gamma^{\dagger}\gamma_0 = (-1)^{h_{\Gamma}} \Gamma \, .
\end{equation}
Thus an operator is permissible only if
\begin{equation}
(-1)^{p_{\bar A}+p_{\Gamma}}=(-1)^{c_{\bar A}+c_{\Gamma}}=(-1)^{h_{\bar
A}+h_{\Gamma}+\epsilon} =1 \, .
\end{equation}

In a next step we will allow for covariant derivatives acting also on
the nucleon fields $\Psi$ and $\bar \Psi$. Since the commutator
between covariant derivatives can be written as
\begin{equation}
\left[ \nabla_{\!\mu},\nabla_{\!\nu}\right] =
\frac{1}{4}\left[u_{\mu}, u_{\nu}\right]
                         - \frac{i}{2}f^{+}_{\mu \nu}-i v^{s}_{\mu \nu}
\end{equation}
any string of derivatives $i\nabla_{\!\mu_1}\dots i\nabla_{\!\mu_n}$ acting
on $\Psi$ can be cast into the form
\begin{equation}
i\nabla_{\!\mu_1}\dots i\nabla_{\!\mu_n}\Psi \rightarrow
\bigl\{i\nabla_{\!\mu_1},\left\{i\nabla_{\!\mu_2},\dots
i\nabla_{\!\mu_n}\right\}\bigr\} \Psi \, ,
\end{equation}
 up to terms with less than $n$   derivatives acting on $\Psi$. The
operators   $\dr$ and $\dl$ are defined as
\begin{align}
\dr =&\,\left\{i\nabla_{\!\mu_1},\left\{i\nabla_{\!\mu_2},\dots
i\nabla_{\!\mu_n}\right\}\right\} \, ,
\\
\dl=&\,
\{i\overleftarrow{\nabla}_{\!\!\mu_1},\{i\overleftarrow{\nabla}_{\!\!\mu_2},
\dots i\overleftarrow{\nabla}_{\!\!\mu_n}\}\}\, ,
\\
 \intertext{where $\overleftarrow{\nabla}_{\!\!\mu}$ is given by}
\overleftarrow{\nabla}_{\!\!\mu } = &\,\overleftarrow{\partial}_{\!\!\mu }
- \left(\Gamma_\mu - i v_\mu^{(s)}\right)\, .
\end{align}
Under charge conjugation, complex conjugation and parity they
transform like
\begin{align}
\left(\psib\dr \Psi\right)^c=&\, \psib\dlr \Psi \, ,
\\
\left( \psib \dr \Psi\right)^\dagger =&\,(-1)^{h_D} \psib\dlr \Psi
                                             \, , \quad h_D=n \, .
\\
\left(\psib\dr \Psi\right)^P=&\,\psib \dru \Psi \, .
\end{align}
The most general term in ${\cal L}_{\pi N}$ with $n$-derivatives on the
nucleon   can take one of the following forms:
\begin{enumerate}
\item[a)] \begin{equation}\label{termone}
  \psib \left(\dl \Gamma \bar A +(-1)^{h_D+h_{\bar A}+h_{\Gamma}} \bar A
\Gamma \drr \right) \Psi
\end{equation}
Charge conjugation invariance, parity and elimination of total derivatives
require
\begin{equation}\label{a}
(-1)^{p_{\bar A}+p_{\Gamma} }
=(-1)^{h_D+c_{\bar A}+c_{\Gamma}}
=(-1)^{h_{\bar A}+h_{\Gamma} }
=1
\end{equation}
\item[b)] \begin{equation}\label{b}
\psib \left(\dl \Gamma i\bar A -(-1)^{h_D+h_{\bar A}+h_{\Gamma}} i\bar A
\Gamma \drr \right) \Psi
\end{equation}
with
\begin{equation}
(-1)^{p_{\bar A}+p_{\Gamma} }
=(-1)^{h_D+c_{\bar A}+c_{\Gamma}}
=-(-1)^{ h_{\bar A}+h_{\Gamma}}
=1
\end{equation}
\end{enumerate}
Note the additional factor $i$ in the term of type b) in \eqref{b}.
Lorentz invariance is again obtained by a suitable contraction of
all Lorentz indices with $g^{\mu\nu}$ and $\epsilon^{\mu\nu\rho\sigma}$.

In the last step we will show that by a suitable redefinition of the nucleon
field $\Psi$ we can eliminate a certain class of operators of the form given
in
the two equations above.
For this purpose we say that an operator of type a) or b)  is of the order
${\cal O}(n,p^m)$, if $n$  derivatives act  on the nucleon fields and if the
chiral
 order of $\bar A$ is $\bar A \backsim {\cal O}(p^m)$. Our strategy is the
following: by a suitable
redefinition of the nucleon fields we can substitute an operator of the
order ${\cal O}(n,p^m)$
by operators ${\cal O}$ with less than $n$-derivatives on the nucleon field,
i.e
${\cal O}\backsim \underset{j<n}{{\cal O}}(j,p^m)$ and by operators
with higher chiral
order, \mbox{${\cal O}\backsim  {\cal O}(n,p^{m+1})$}. The successive
redefinitions of the nucleon fields   eliminates   this operator up to terms
with no derivatives on $\Psi$, which can be absorbed in the existing
lagrangian
$\cL_{\pi N}$, and terms with higher chiral order.\\
Applying the    nucleon field transformation
\begin{equation}
\Psi =
\left[ 1+ \bar A \Gamma \overrightarrow{D}^{\! n-1}_{\!\!\mu_{2} \dots
\mu_n} \right]\Psi'\, \quad \bar A \backsim {\cal O}(p^m) \, ,\,\,\Gamma\,
\in\text{Clifford Algebra}\label{transform}
\end{equation}
to the leading order relativistic lagrangian  $\cL_{\pi N}^{(1)}$ in
\eqref{LMB1}
 generates the following lagrangian up to irrelevant fore-factors
\begin{equation}
\begin{split}
{\cal L }_{\text{ind}} &\, =
\psib'\left[ \dl \Gamma \gamma^{\mu_1} \bar A +
(-1)^{h_{\bar A} +h_{\Gamma}+h_D} \bar A \gamma^{\mu_1} \Gamma
\dr\right]\Psi'
\\
&\,+\underset{j<n}{O}(j,p^m)+\underset{l\geqslant n}{O}(l,p^{m+1})
\\
\end{split}
\end{equation}
The products $\Gamma \gamma^{\mu_1}$ and $ \gamma^{\mu_1}\Gamma$  can be
reduced to   elements of the Clifford algebra and one obtains operators of
the type a) or b).   \\
 As an example we choose
$\Gamma = \gamma^{\mu}$. In this case the induced lagrangian is found to be
\begin{equation}
\begin{split}
{\cal L }^{\gamma^{\mu}}_{\text{ind}} &\, =
\psib'\left[ \dl  g^{\mu \mu_1} \bar A +
(-1)^{h_{\bar A} +h_D} \bar A g^{\mu \mu_1}   \dr\right]\Psi'
\\
&\,+  \psib'\left[ \dl \Gamma \sigma^{\mu \mu_1}i\bar A -
(-1)^{h_{\bar A} +h_D} i\bar A \sigma^{\mu \mu_1}   \dr\right]\Psi'.
\\
\end{split}
\end{equation}
The first line is of type a) and the second is of type b). It is also
possible to obtain the reverse situation by substituting $\Gamma \to
i\Gamma$ in equation \eqref{transform}.
Notice that there is a sign difference between the two  terms on each line.
This means that one of these lines can be eliminated via partial integration
up
to operators with higher chiral order than $\bar A $. The remaining line can
then be eliminated via the field redefinition \eqref{transform}.
Letting $\Gamma$ run through all elements of the Clifford Algebra one
similarly finds
\begin{enumerate}
\item \begin{equation} \psib \bar A \,\Gamma^{\mu \mu_1}   \dr \Psi +h.c.
       \backsimeq 0 \, ,\label{one}\end{equation}
where $\Gamma^{\mu \mu_1}=\sigma^{\mu \mu_1},\gamma_5\sigma^{\mu \mu_1}$
\item \begin{equation} \psib \bar A \, \Gamma  \dr \Psi +h.c.
       \backsimeq 0 \, ,\label{two}\end{equation}
where $ \Gamma= {\bf 1}, \gamma_5$
\item \begin{equation} \psib \bar A \, \Gamma^{\nu}
 \overrightarrow{D}^{\! n}_{\!\!\mu_1 \mu_2\dots \mu_n} \Psi
       \backsimeq  \, \psib \bar A \, \Gamma^{\mu_1}
\overrightarrow{D}^{\! n}_{\!\!\nu\mu_2 \dots \mu_n} \Psi, \end{equation}
where $\Gamma^{\mu}=\gamma^{\mu},\gamma_5\gamma^{\mu}$
\item \begin{equation} \psib \bar A \,\epsilon^{\alpha \beta \mu \mu_1}\,
        \Gamma_{\mu}   \dr \Psi +h.c. \backsimeq 0 \end{equation}
\item \begin{equation} \psib \bar A \,\epsilon^{\alpha \beta \mu \mu_1}\,
        \Gamma_{\mu \lambda}   \dr \Psi +h.c. \backsimeq 0 
\end{equation}
\item \begin{equation} \psib \bar A \, \Gamma^{\mu_1}  \dr \Psi +h.c.
       \backsimeq 0 \, . \end{equation}
\end{enumerate}
Here ``$\backsimeq$'' stands for ``equivalent to terms with less derivatives
on
$\Psi$ and higher order terms''. It should be noted, however, that our
elimination procedure only works, if the chiral order of $\bar A$ is at
least    $\bar A \sim {\cal O}(p)$.
In \cite{Kr90} the equations of motion have been used in order to
remove six equation-of-motion type operators in the $SU(3)$
relativistic lagrangian. The first five operators are the $SU(3)$
versions  of the respective terms in the above list. However,
the last operator in \cite{Kr90} does not appear in our list.

Finally we indicate those terms with derivatives on the nucleon fields 
which cannot be eliminated by applying the rules given above.
Let ${\bar A}^{\nu_1 \dots \nu_m}$ be a chiral operator of the form
\eqref{operator}, where all Lorentz indices are fully contracted except for the
indices $\nu_1 \dots \nu_m$. It is  understood that no two of these
indices are due to the same metric tensor $g^{\nu_i \nu_j}$ and that no two or
more of the indices are due to the same antisymmetric
tensor $\epsilon$. The most general terms with $n$-derivatives on the
nucleon field of the form \eqref{termone} which cannot be eliminated 
by nucleon field redefinitions are then
\begin{align}
 \psib \left(\dl \Gamma_{\rho_1} {\bar A}^{\rho_1 \mu_1 \dots \mu_n}
 +(-1)^{h_D+h_{\bar A}+h_{\Gamma}} {\bar A}^{\rho_1 \mu_1 \dots \mu_n}
\Gamma_{\rho_1} \drr \right) \Psi \, \\
\intertext{with
$\Gamma_{\rho_1} = \gamma_{\rho_1},\gamma^5\gamma_{\rho_1}$ and }
\psib\left(\dl \Gamma_{\rho_1\rho_2}
{\bar A}^{\rho_1\rho_2\mu_1\dots\mu_n}
 +(-1)^{h_D+h_{\bar A}+h_{\Gamma}} {\bar A}^{\rho_1\rho_2\mu_1\dots\mu_n}
\Gamma_{\rho_1\rho_2} \drr \right) \Psi \,
\intertext{with
$\Gamma_{\rho_1\rho_2} =
\sigma_{\rho_1\rho_2},\gamma^5\sigma_{\rho_1\rho_2}$.}
\nonumber\end{align}
An analogous expression holds for terms of the form \eqref{b}.
Consequently, if $n$ derivatives act on the nucleon field the chiral order of
$\bar A$
must be of the order $\bar A \sim {\cal O}(p^{n+1})$ for 
$\Gamma =\Gamma^{\rho_1}$ and of the order $\bar A \sim {\cal
O}(p^{n+2})$ for $\Gamma =\Gamma^{\rho_1 \rho_2}$. If
one of the indices $\nu_j$ in $A^{\nu_1 \dots \nu_m}\,$
is due to the antisymmetric tensor $\epsilon$ the chiral order is
further increased by at least one unit. These observations restrict severely
the possible terms with $n$ derivatives on the nucleon field of a given 
chiral order.

 \begin{table}[h]
\renewcommand{\arraystretch}{1.6}
\begin{center}
\begin{tabular}{|c|c|c|c|}
\hline
                &      C              & P               &   h.c.         \\
\hline
$u_\mu         $&$u^T_\mu          $&$-u^\mu          $&$ u_\mu         $\\
$f^{+}_{\mu\nu}$&$-f^{+T }_{\mu\nu}$&$f^{\mu\nu}_{+ } $&$f^{+ }_{\mu\nu}$\\
$f^{-}_{\mu\nu}$&$f^{-T }_{\mu\nu} $&$-f^{\mu\nu}_{- }$&$f^{- }_{\mu\nu}$\\
$v^{s}_{\mu\nu}$&$-v^{sT}_{\mu\nu} $&$v^{\mu\nu}_{s}  $&$v^{s}_{\mu\nu} $\\
$\chi_+        $&$\chip^T          $&$\chip           $&$\chip          $\\
$\chi_-        $&$\chim^T          $&$-\chim          $&$-\chim         $\\
$\overrightarrow{\nabla}_{\!\!\!\mu}$&$\overleftarrow{\nabla}^T_{\!\!\mu}
       $&$\overrightarrow{\nabla}^{\mu}$&$
\overleftarrow{\nabla}_{\!\!\mu}$\\
\hline
\end{tabular}
\end{center}
\caption{P, C, and h.c. transformation properties of the chiral fields.}
\label{chiralc}
\end{table}

\begin{table}
\newcommand{\gfmu}{\gamma^5\gamma^{\mu }}
\newcommand{\sigmn}{\sigma^{\mu \nu}}
\newcommand{\gmz}{\gamma^0}
\renewcommand{\arraystretch}{1.6}
\begin{center}
\begin{tabular}{|c|c||c||c|c|}
\hline
            &      C                 &       h.c                     &
           &      P       \\
${\boldsymbol{1_d}} $&${\boldsymbol{1_d}} $&$\gmz \,{\boldsymbol{1_d}}
\,\gmz=       {\boldsymbol{1_d}}$&$
\psib  \Psi         $&$\psib  \Psi        $
\\
$\gamma^\mu            $&$-\gamma^{\mu T}        $&$\gmz \gamma^\mu \gmz =
\gamma^{\mu \dagger}$&$
\psib \gamma_\mu  \Psi $&$\psib \gamma^\mu  \Psi$
\\
$\gamma^5          $&$\gamma^{5T}          $&$\gmz \gamma^5 \gmz
=-\gamma^{5\dagger}        $&$
\psib  \gamma^5\Psi$&$-\psib \gamma^5 \Psi $
\\
$\gfmu           $&$\left(\gfmu\right)^T  $&$\gmz \gfmu \gmz
=\left(\gfmu\right)^{\dagger} $&$
\psib  \gamma^5 \gamma_\mu\Psi$&$-\psib \gamma^5 \gamma^\mu  \Psi $
\\
$\sigmn    $&$-\left(\sigmn\right)^T$&$\gmz \,\sigmn \gmz
=\left(\sigmn\right)^{\dagger}$&$
\psib \sigma_{\mu \nu} \Psi$&$\psib \sigma^{\mu \nu} \Psi $\\
\hline
\end{tabular}
\end{center}
\caption{P, C, and h.c. transformation properties of the elements of the
Clifford Algebra.}
\label{cliffc}
\end{table}
\clearpage

\section{Loop-Functions}
\label{appendixc}
We collect the loop integrals employed in this
article. The following definitions and results have been used
\begin{equation}
\Delta (M^{2}_{\pi})=  \frac{1}{i} \int \frac{d^d  k}{(2\pi)^d}
                                \frac{1}{\mpi-k^2}
=2 \mpi \left[ \, \Lambda (\mu) + \frac{1}{32 \pi^2} \ln\frac{\mpi}{\mu^2}
\right] \, ,
\end{equation}
with
\begin{equation}
\Lambda (\mu)  =  \frac{\mu^{d-4}}{16 \pi^2} \left\{
\frac{1}{d-4}-\frac{1}{2}
               \left[ \, \ln(4\pi)+\Gamma'(1)+1 \right]  \right\} \, .
\end{equation}
The integral $J_{\pi\pi}$ is defined as
\begin{equation}
J_{\pi\pi}( p^2)=
                            \frac{1}{i} \int\frac{d^d  k}{(2\pi)^d}
                                \frac{1}{\mpi-k^2} \frac{1}{\mpi-(k-p)^2}
\end{equation}
with explicit representation
\begin{equation}
\begin{split}
 J_{\pi\pi}(p^2)= &\, \bar J_{\pi\pi}(p^2)+ J_{\pi\pi}(0)
\\
J_{\pi\pi}(0) =& \,  -2\Lambda (\mu)
-\frac{2}{32 \pi^2}\left[ \ln\frac{M^{2}_{\pi}}{\mu^2} +1\right]+{\cal
O}(d-4)  
\\
r=&\,\left|1-4\frac{\mpi}{p^2}\right|^{\frac{1}{2}} \\[0.3cm]
\bar J_{\pi\pi}(p^2)= &\,
\begin{cases}
\frac{1}{16\pi^2}\left[ \, 2 -2r\arctan \frac{1}{r} \right]
\, ,&\,
0<p^2<4\mpi
\\ [0.5cm]
\frac{1}{16\pi^2}\left[ \, 2-r\ln\left|\frac{1+r}{1-r}\right| +i\pi r
\right]
\, ,&\, p^2>4\mpi \, .
\end{cases}
\end{split}
\end{equation}
The integral $K_0$ is defined as
\begin{equation}
K_0(\omega,p^2)=
                            \frac{1}{i} \int\frac{d^d  k}{(2\pi)^d}
                                \frac{1}{\mpi-k^2} \frac{1}{\mpi-(k-p)^2}
                                \frac{1}{v\cdot k-\omega } \, ,
\end{equation}
and we find
\begin{equation}
K_0(0,p^2) =
\begin{cases}
\frac 1{16\pi \sqrt{p^2}}\ln \frac{2M_{\pi}-\sqrt{p^2}}{2M_{\pi}+\sqrt{p^2}}
\, , &
\omega=0 \text{\quad and\quad }  0<p^2<4\mpi
\\ [0.5cm]
\frac 1{16\pi \sqrt{p^2}}
\left[ \ln \frac{\sqrt{p^2}-2M_{\pi}}{\sqrt{p^2}+2M_{\pi}}-i\pi \right]
\, , &
\omega=0      \text{\quad and\quad }   p^2>4\mpi  \, .
\end{cases}
 \end{equation}
Finally, we have defined $J_{0}(\omega)$ as
\begin{equation}
\begin{split}
J_{0}(\omega)= &\,\frac{1}{i} \int\frac{d^d  k}{(2\pi)^d}
                                \frac{1}{\mpi-k^2}\frac{1}{v\cdot k-\omega }
\\
= &\,
-4\Lambda (\mu)\, \omega
+{\omega \over 8\pi^2} \left[ 1-\ln{{\mpi\over\mu^2}}\right]
-\frac{1}{4\pi^2} \sqrt{\mpi-\omega^2}\arccos{{-\omega\over M_{\pi}}} \, .
\end{split}
\end{equation}
The derivative of $J_{0}(\omega)$ with respect to $\omega$ at $\omega =0$ is
given by
\begin{equation}
\begin{split}
J^{'}_{0}(0)= &\,\frac{1}{i} \int\frac{d^d  k}{(2\pi)^d}
                                \frac{1}{\mpi-k^2}\frac{1}{(v\cdot k)^2 }
 =
-4\Lambda (\mu) -\frac{1}{8\pi^2}\left[1+\ln{\frac{\mpi}{\mu^2}}\right] \, .
\end{split}
\end{equation}
\newpage

\end{document}